\documentclass[12pt]{article}
    \usepackage{times}
\usepackage{srcltx}
\expandafter\let\csname equation*\endcsname\relax
\expandafter\let\csname endequation*\endcsname\relax

\usepackage[mathscr]{eucal}
\usepackage{amsmath,amsfonts,amssymb,amsthm}
\usepackage{stmaryrd}
\usepackage{hyperref}
\newcommand{\changed}[1]{{#1}}

\def\sd{S^\dagger}
\def\bsd{\bar S^\dagger}
\def\bs{\bar S}

\def\amb{\fR^{d,2}\backslash\{0\}}

\def\BGST{Barnich:2004cr}
\def\BGadS{Barnich:2006pc}

\def\BGnlp{Barnich:2010sw}
\def\BG-Poincare{Barnich:2009jy}
\def\Fedosov-book{Fedosov:1996fu}

\advance\textheight\topskip
\numberwithin{equation}{section} \makeatletter
\@addtoreset{equation}{section}

\renewcommand{\tilde}{\widetilde}
\renewcommand{\hat}{\widehat}
\newtheorem{prop}{Proposition}%[section]
\newtheorem{lemma}[prop]{Lemma}

\renewcommand{\simeq}{\cong}

\newcommand{\bref}[1]{\textbf{\ref{#1}}}

\newcommand{\gh}[1]{\mathrm{gh}(#1)}

\newcommand{\assalgebra}{\mathscr}    %ass[ociative] algebra(s)
\newcommand{\algA}{\assalgA}
\newcommand{\assalgA}{\assalgebra{A}}

\newcommand{\dd}{\partial}
\renewcommand{\d}{\partial}

\newcommand{\tensor}{\otimes}

\renewcommand{\geq}{\,{\geqslant}\,}
\renewcommand{\leq}{\,{\leqslant}\,}

\newcommand{\inner}[2]{\langle #1{,}\,#2\rangle}
\newcommand{\binner}[2]{%
  {\langle}\kern-4.15pt{\langle}#1{,}\,#2{\rangle}\kern-4.15pt{\rangle}}
\newcommand{\commut}[2]{[#1{,}\,#2]}
\newcommand{\qcommut}[2]{[#1{,}\,#2]_*}
\newcommand{\pb}[2]{\left\{{}#1{},{}#2{}\right\}}

\newcommand{\half}{\mathchoice{%
    \ffrac{1}{2}}{\frac{1}{2}}{\frac{1}{2}}{\frac{1}{2}}}

\newcommand{\ffrac}[2]{\raisebox{.5pt}%
  {\footnotesize$\displaystyle\frac{#1}{#2}$}\kern1pt}

\newcommand{\brst}{\mathsf{\Omega}}

\newcommand{\dl}[1]{\mathchoice{\ffrac{\dd}{\dd #1}}{\frac{\dd}{\dd
      #1}}{\ffrac{\dd}{\dd #1}}{\ffrac{\dd}{\dd #1}}}

\def\const{\mathop\mathrm{const}\nolimits}

\newcommand{\fR}{\mathbb{R}}

\newcommand{\derham}{\boldsymbol{d}}

\newcommand{\manifold}[1]{\mathscr{#1}}
\newcommand{\manX}{\manifold{X}}

\def\cD{\mathcal{D}}

\def\cH{\mathcal{H}}

\numberwithin{equation}{section} \makeatletter
\@addtoreset{equation}{section}

\begin{document}

%\begin{titlepage}
\begin{center}
\textbf{\Large{
Higher-order singletons, partially massless fields,\\ and their boundary values in\\[5pt]
the ambient approach}}

\vspace{1.4cm}

  {\large {Xavier Bekaert$^a$ and Maxim Grigoriev$^b$}}
  
  \end{center}
%
%
%\end{titlepage}
%\maketitle

%\pagebreak

%\address
\begin{center}
{$^a$~Laboratoire de Math\'ematiques et Physique Th\'eorique\\
    Unit\'e Mixte de Recherche 7350 du CNRS\\
    F\'ed\'eration de Recherche $2964$ Denis Poisson\\
    Universit\'e Fran\c cois Rabelais,
    Parc de Grandmont, 37200 Tours, France}\\[10pt]
%\address
{$^b$~Tamm Theory Department,
 Lebedev Physics Institute\\
 Leninsky prospect 53, 119991 Moscow, Russia}
\end{center}
\vspace{1cm}
\begin{abstract}
Using ambient space we develop a fully gauge and $\mathfrak{o}(d,2)$ covariant approach to boundary values of $AdS_{d+1}$ gauge fields. It is applied to the study of (partially) massless fields in the bulk and (higher-order) conformal scalars, i.e. singletons, as well as (higher-depth) conformal gauge fields on the boundary. In particular, we identify the corresponding {generalized} Fradkin--Tseytlin equations as obstructions to the extension of the off-shell boundary value to the bulk, generalizing the usual considerations for the holographic anomalies 
to the partially massless fields. We also relate the background fields for the higher-order singleton to the boundary values of partially massless fields and prove the appropriate generalization of the Flato--Fronsdal theorem,
which is in agreement with the known structure of symmetries for the higher-order wave operator.
All these facts support the following generalization of the higher-spin holographic duality: the $O(N)$ model at a multicritical isotropic Lifshitz point should be dual to the theory of partially massless symmetric tensor fields described by the Vasiliev equations based on the higher-order singleton symmetry algebra.
\end{abstract}
\thispagestyle{empty}

\newpage
\tableofcontents%[hideallsubsubsections]

\section{Introduction}

In agreement with the holographic picture, most of the structures underlying totally symmetric higher-spin (HS) gauge fields around an anti de Sitter (AdS) background are determined by those of a scalar singleton -- the conformal scalar field living on the conformal boundary \cite{Flato:1978qz,Konstein:2000bi,Sezgin:2002rt}.
From the group-theoretical point of view, this relation is known as the Flato--Fronsdal theorem that states that the tensor product of two free singletons decomposes into the direct sum of conformal conserved currents \cite{Flato:1978qz,Vasiliev:2004cm}, which may be identified with certain boundary values of free AdS HS gauge fields. Furthermore, the bosonic higher-spin algebra
can be seen as the algebra of higher symmetries for the free singleton \cite{Eastwood:2002su,Vasiliev:2003ev}, while the background fields (also called shadow fields in this context) coupling minimally to the singleton are identified with the appropriate boundary values of the bulk on-shell HS gauge fields.

The present paper is devoted to a natural generalization of this picture, where the scalar singleton is replaced by a higher-order one, \textit{i.e.} a boundary scalar field $\phi_0$ with appropriate conformal weight %$\Delta=\frac{d}2-\ell$
satisfying the conformally-symmetric higher-order wave equation (called, more generally, the polywave equation of order $2\ell$) \begin{equation}
(\Box_0)^\ell \phi_0=0,\label{polywave}
\end{equation}
where $\Box_0$ denotes the d'Alembertian on Minkowski spacetime and $\ell$ is a positive integer.
In particular, the tensor product of two such higher-order singletons decomposes into the direct sum of certain massless \cite{Fronsdal:1978vb} and partially massless \cite{Deser:1983tm,Deser:1983mm,Deser:2001pe} symmetric tensor fields
in the bulk. This provides an interpretation of PM fields of odd depths as ``doubletons'', \textit{i.e.} composite fields made of two singletons. In the same spirit, one identifies the higher-order analogues of the HS algebra by studying
the algebra of higher symmetries for the equation \eqref{polywave}. These algebras are known in the mathematical literature~\cite{Eastwood:2005,Gover:2009}. It turns out that the nontrivial symmetries of higher-order singletons are in one-to-one correspondence with (from the bulk perspective) the vacuum symmetries of partially massless fields and with (from the boundary perspective) the higher-order conformal Killing tensors which are the global reducibility parameters of higher-depth shadow fields.
As a rule, the identification of non-Abelian HS algebras
is important because it suggests that the picture holding at free level may be extended to full non-linear level.
In the present case, the Vasiliev equations of bosonic HS gravity in any dimension \cite{Vasiliev:2003ev} indeed remain consistent
if one quotients the off-shell HS algebra by the smaller ideal corresponding to higher-order singletons. That the off-shell HS algebra (and  some of its quotient) should somehow correspond to a HS gravity with partially massless
fields in the spectrum seems to be common lore for experts on the subject (see \textit{e.g.} the brief remarks
in \cite{Bekaert:2005vh, Alkalaev:2007bq}).
%However, we are not aware of any previous precise identification of the suitable non-trivial quotient and %corresponding spectrum in the literature.}

From a quantum field theory (QFT) perspective, AdS higher-spin theories might 
provide semiclassical bulk duals of some celebrated QFTs. For instance, the 
critical $O(N)$ model at the Wilson-Fisher (or Gaussian) fixed point of the 
renormalization group (RG) has been conjectured to admit a holographic dual 
description as a HS field theory with (un)broken gauge symmetries 
\cite{Sezgin:2002rt}. Our present generalization of HS 
holography to higher-order singletons can be motivated by the existence of 
special RG fixed points: the multicritical isotropic Lifshitz\footnote{Although 
the union of the adjectives ``isotropic'' and ``Lifshitz'' may sound like an 
oxymoron (since, usually, the celebrated ``Lifshitz point'' rather correspond to 
some anisotropic scale symmetry), we followed the standard RG terminology 
{for these} multicritical fixed points (see \textit{e.g.} \cite{Babich}). In 
most physical applications, people focus on the fourth order ($\ell=2$) in the 
derivative expansion (\textit{c.f.} \cite{Diehl} for a review).} points, which 
can be described precisely by scalar fields with a polywave kinetic operator. 
The existence of these special fixed points prompts a natural generalization of 
the conjectures \cite{Sezgin:2002rt} that will be spell out in 
details in the conclusion.

In order to describe higher-order singletons, partially massless fields, their 
symmetries and their boundary values, we make use and develop a manifestly 
local, coordinate-independent, gauge- and $\mathfrak{o}(d,2)$-covariant approach to the boundary values of 
AdS gauge fields. In contrast to usual approaches, it does not rely on gauge 
fixations and hence allows to trace the gauge invariance at all stages of 
the procedure (note, however, the approach of~\cite{Metsaev:2008fs}
using only partial gauges). In this respect, we also slighly improve our 
previous method~\cite{Bekaert:2012vt} where certain partial gauges where 
utilized. From a technical point of view, we extensively employ the ambient space 
along with the first-quantized description of free gauge fields and its {parent BRST
generalization~\cite{Barnich:2004cr,Barnich:2006pc,Alkalaev:2009vm,Bekaert:2009fg}}. One should mention also a certain similarity with the conformal geometry methods~\cite{FG,Eastwood:2002su} and the unfolded formalism~\cite{Vasiliev:1988xc,Shaynkman:2004vu,Vasiliev:2009ck,Vasiliev:2012vf,Didenko:2012vh}.

An additional concept which appears useful in the context of boundary values is that of background fields.
Given a free gauge theory (seen either as a theory of free classical gauge fields or as a constrained system describing the respective first-quantized particles) one can associate to it a set of background fields minimally coupled to the free system~\cite{Segal:2002gd,Vasiliev:2005zu,Grigoriev:2006tt,Bekaert:2009ud}. The nonlinear off-shell constraints and gauge symmetries for background fields are determined by the free system. In this way, the nonlinear structure of a dual bulk theory is to some extent determined by the boundary system. More precisely, the HS algebra in the bulk coincides with the gauged algebra for the background fields which is in turn the maximal algebra of global symmetries of the boundary system. At the same time, linearized background fields for the conformal scalar can be identified with the boundary values of the bulk HS fields. We present evidences for the extension of such correspondences to the case of higher-order singleton and the associated HS multiplet of bulk (
partially) massless fields.

\bigskip

%The plan of the paper is as follows. 
We begin this paper with a discussion of (higher-order) singletons as on-shell fields on various spaces (ambient, bulk and boundary) which allows to highlight and to give a flavor of some points that apply in a similar way to (partially) massless fields, but avoiding technical complications.
More precisely, Section \bref{singletons} contains a quick review of the standard asymptotic solutions of Klein-Gordon equation, focusing on the critical mass case corresponding to (higher-order) singletons.  The leading boundary value must be constrained to be on-shell if one wants to avoid the introduction of logarithmic term in the expansion. As further discussed, this latter requirement appears to be very natural from the ambient point of view.
The section \bref{preliminaries} provides a group-theoretical catalogue
of the $\mathfrak{o}(d+2)$-modules that are relevant for the present paper and ends with the statement of our generalization of the Flato--Fronsdal theorem.
In Section \bref{Partiallymassless}, the ambient and parent formulations of PM fields are reviewed
in order to determine their global reducibility parameters. A reducible multiplet of PM fields is introduced which fits the Flato--Fronsdal theorem in a natural way.
The gauge and $\mathfrak{o}(d,2)$ covariant approach to the boundary data for PM fields is addressed in Section \bref{Bdyvalues} while the manifestly conformal formulations of higher-order singletons, higher-depth Fradkin--Tseytlin and shadow fields, partially conserved currents and higher-depth conformal Killing tensors are presented in Section \bref{sec:manifest}.
Higher-order singletons, their background fields and their symmetries are the subjects of Section \bref{Backgroundfields}.
The conclusion provides a summary of our main results.
Finally, some technical proofs and some toy model examples have been placed in Appendix.

\section{Higher-order singletons: boundary, bulk, and ambient descriptions}\label{singletons}

The subsection \bref{intrinsic} contains a review of the standard AdS/CFT correspondence for a scalar field (see \textit{e.g.}~\cite{Skenderis:2002wp}) where we focus on the particular case corresponding to the (higher-order) singleton and discuss some of its subtleties. 
The subsection \bref{subsambound} is a considerably expanded review of the ambient approach to the boundary values of bulk scalar fields\footnote{The ambient space approach for the metric tensor \textit{\`a la} Fefferman-Graham has been applied since the early days of the AdS/CFT correspondence (see \textit{e.g.} the review \cite{Alvarez:2002di} and references therein).} that was shortly presented in \cite{Bekaert:2012vt} and that we present with more explicit details, explanations and discussions of the possible choices of ambient lift.

\subsection{Boundary values in terms of intrinsic AdS geometry}\label{intrinsic}

Let $\varphi$ be an AdS scalar of mass $m$ satisfying Klein--Gordon equation 
\begin{equation}
\label{ads-scalar}
 (\nabla^2-m^2)\varphi=0\,,
\end{equation}
where $\nabla^2$ is the Laplace-Beltrami operator on $AdS_{d+1}$.
We work in the coordinates ($x^a,\rho$) on $AdS_{d+1}$ such that the metric has the form
\begin{equation}
\label{coord}
 ds^2=\frac{1}{4\rho^2}(d\rho)^2+\frac{1}{\rho}(\eta_{ab}dx^adx^b)\,, \qquad (a,b=0,\ldots, d-1)
\end{equation}
and the conformal boundary is located at $\rho=0$. 

There are two possible $\mathfrak{o}(d,2)$-invariant choices of asymptotic behavior for the on-shell scalar field:
\begin{equation}
\label{anz}
 \varphi(x,\rho)=\rho^{\Delta_{\pm}/2}\phi_\pm(x,\rho)\,,
\end{equation}
where $\phi_\pm(x,\rho)$ is regular at $\rho=0$. These two choices correspond to the two solutions $\Delta_{\pm}$ (with $\Delta_-\leqslant\Delta_+$) of the algebraic equation
\begin{equation}
\label{anz2}
m^2=\Delta(\Delta-d)\quad\Leftrightarrow\quad \Delta_\pm=\half(d\pm\sqrt{d^2+4m^2})\,.
\end{equation}
The $\mathfrak{o}(d,2)$ invariance of the condition means that $\phi^\pm_0(x)=\phi_\pm(x,0)$ transforms through itself, \textit{i.e.} determines
a conformal field on the boundary (of weight $\Delta_\pm$). This conformal field $\phi_0$ is usually refered to as the boundary value of the on-shell scalar field $\varphi$ associated to the asymptotic behavior $\Delta_{\pm}$.
The lowest root $\Delta_-$ corresponds to the leading boundary behaviour while the highest root $\Delta_+$ describes the subleading boundary behaviour (since $\Delta_-\leqslant\Delta_+$).

Due to the ansatz \eqref{anz} where $\Delta_{\pm}$ is \changed{a} solution of \eqref{anz2}, the Klein--Gordon equation \eqref{ads-scalar} reads in terms of the function $\phi_\pm(x,\rho)$ in \eqref{anz} as follows:
\begin{equation} \label{feq2}
\Box_0 \phi_\pm + 2 (d - 2 \Delta_\mp + 2+ 2 \rho \partial_\rho) \partial_\rho \phi_\pm 
=0\,,
\end{equation}
where $\Box_0=\eta^{ab}\dl{x^a}\dl{x^b}$.
Usually, one looks for solutions $\phi_\pm(x,\rho)$ admitting some asymptotic expansion (as Taylor or Frobenius series) in the radial coordinate $\rho$. For instance, if $\phi_\pm$ admits a power series expansion
\begin{equation}\label{powerexp}
 \phi_\pm(x,\rho)=\phi_0(x)+\rho\,\phi_1(x)+\rho^2\phi_2(x)+\ldots  \,,
\end{equation}
then its coefficients $\phi_n(x)$ satisfy the recurrence relations
\begin{equation}\label{recurrence}
2n\, (2 \Delta_\mp - d - 2n)\phi_{n} = \Box_0 \phi_{n-1}\,,\quad(n=1,2,\ldots)\,,
\end{equation}
in order for \eqref{powerexp} to be solution of \eqref{feq2}.
As long as $2 \Delta_\mp - d \neq 2n$, this recurrence relation can be solved without any subtlety. 
For instance, the subleading asymptotic solution $\phi_+$ (corresponding to the highest root $\Delta_+$) always admits an asymptotic expansion as a power series {(since $2 \Delta_- - d\leq 0$)}.

However, if there exists a positive integer $\ell$ such that $2 \Delta_+ - d = 2\ell$, then the
recurrence relation \eqref{recurrence} for the power series expansion of the leading asymptotic solution $\phi_-$  (corresponding to the lowest root $\Delta_-$) 
meets in general an obstruction at $n=\ell$. The roots corresponding such an obstruction are thus: $\Delta_\pm=\frac{d}2\pm \ell$. 
The corresponding AdS scalar field $\varphi$ has mass-square $m^2=\ell^2-(\frac{d}2)^2$ and will be called ``critical''.
Two natural options arise to remove the obstruction: The standard procedure (see \textit{e.g.}~\cite{Skenderis:2002wp}) is to replace the Taylor series \eqref{powerexp} by a Frobenius series via the introduction of a logarithmic term:
\begin{equation}\label{powelogrexp}
 \phi_-(x,\rho)=\phi_0(x)+\rho\,\phi_1(x)+\ldots +\rho^{d/2-\Delta_-
} \big(\,\log \rho\, \phi_\ell(x)\,+\,\psi_\ell(x)\,\big)+\,\ldots\,,
\end{equation}
More precisely the solution reads as
\begin{eqnarray}\label{log}
 \varphi(x,\rho)&=&\rho^{\Delta_-/2}\Big(\,\phi_0(x)+\rho\,\phi_1(x)+\ldots +\rho^{d/2-\Delta_-
} \log \rho\, \phi_\ell(x)+\ldots\,\Big)\nonumber\\
&&+\rho^{\Delta_+/2}\Big(\,\psi_\ell(x)+\rho\,\psi_{\ell+1}(x)+\ldots\,\Big)\,.
\end{eqnarray}
so that $\psi_\ell(x)$ can be interpreted as the boundary value of the subleading solution.

An alternative is to keep a power series expansion but to impose the consequence of the relations \eqref{recurrence} for the leading boundary value $\phi_0$,  \textit{i.e.} to impose the polywave equation \eqref{polywave}. 
In this case the term with the logarithm is missing in~\eqref{log} while the subleading boundary value $\psi_\ell(x)$ is unconstrained as before.
Thus, as a boundary field, the higher-order singleton is described by an on-shell scalar field $\phi_0$ 
while, as a bulk field, the higher-order singleton is represented by the modes $\varphi$ of a critical scalar field 
with $\phi_\ell=0$ and quotiented by the subleading solutions given by the second line of~\eqref{log} i.e. solutions of the form $\rho^{\Delta_+/2}\psi(x,\rho)$ where the function $\psi(x,\rho)$ is regular at $\rho=0$.

As one can see, the bulk equation of motion \eqref{ads-scalar} for a  scalar field with critical mass can impose conditions on the leading boundary value $\phi_0$ depending on the precise requirements imposed on the class of asymptotic expansion for the function $\phi_\pm$ (and therefore for the AdS scalar field $\varphi$). For instance, requiring $\phi_-$ to be a smooth function in $\rho$ on the interval $0\leq \rho<\epsilon$ (here a small $\epsilon$ is chosen to restrict to near-boundary analysis)  imposes the polywave equation on $\phi_0$ if $d-2\Delta=2\ell$ with $\ell$ a positive integer. Note that it is actually more convenient to work in the interval $-\epsilon < \rho <\epsilon$ because it simply corresponds to a neigbourhood of a hypercone in the ambient space (as explained in the next subsection). If instead one requires $\phi_-$ to be smooth only for $\rho>0$ and to have a limit at $\rho=0$ the leading boundary value becomes unconstrained because in this space one can always add a term 
$\rho^\ell \phi_\ell \log \rho$ to relax the condition. This term is not smooth at $\rho=0$ (since its $\ell$-th derivative contains a term proportional to $\log \rho\, \phi_\ell$).

Another natural question is to which extent the solution is entirely determined by $\phi_0$. This again highly depends
on the precise class of functions to which $\phi_-$ belongs to. If one does not require smoothness at $\rho=0$ one can clearly add to a given solution a one with a subleading asymptotic so that it explicitly reads as in \eqref{log} (the term with $\log$ is only present if $d/2-\Delta_-$ is a positive integer).
The bulk equations of motion leave $\psi_\ell(x)$ undetermined. In this case the near-boundary solution is determined by two independent boundary fields $\phi_0$ and $\psi_\ell$. 
%It is natural to refer to $\phi_0,\psi_\ell$ as boundary data for the bulk field $\phi$.
If instead one requires $\phi_-(x,\rho)$ to be smooth at $\rho=0$ one can only add a subleading term if $d/2-\Delta_-$
is positive integer because not all derivatives with respect to $\rho$ of $\rho^{d/2-\Delta_-}$ are defined at $\rho=0$. 
Moreover, as we have seen, in this case $\log \rho$ is also not allowed 
and this subjects $\phi_0$ to the conformal equation $\Box_0^\ell\phi_0=0$ while $\psi_l$ remains unconstrained. 
To summarize: for $\phi_-$ smooth at $\rho=0$, the boundary data is given by only 
an unconstrained $\phi_0(x)$ if ${d-2\Delta_-}\neq 2\ell,\,\, \ell>0$, or by 
a conformal scalar $\phi_0$ subject to $\Box_0^\ell\phi_0=0$ and an unconstrained $\psi_\ell$ if ${d-2\Delta_-}=2\ell,\,\, \ell>0$.

Requiring $\phi_\pm(x,\rho)$ to be smooth around $\rho=0$ can look somewhat 
unnatural from the AdS point of view as the boundary $\rho=0$ does not belong to 
$AdS$. However, this choice is  rather natural from the ambient space perspective (reviewed in the next 
subsection) where both the AdS field and its boundary value are represented by 
an ambient space field. Moreover, smoothness of all fields is a crucial 
assumption for the HS fields at the interaction level because the nonlinear theory of HS gauge fields
involves derivatives of infinite order. Even at the linear level, the assumption is also natural
in the unfolded formulation (or jet space) framework.

In contrast to the above conditions affecting near-boundary behavior only, another alternative is to restrict
the behavior of $\phi$ in the AdS interior. The important choice employed in the literature is to require
$\phi$ to be regular in the deep interior. 
This is known to uniquely determine $\psi_\ell$ in terms of $\phi_0$ so that
the boundary data again reduces to just $\phi_0$. Moreover, according to the conventional AdS/CFT prescription,
correlation functions of the boundary conformal operators are encoded in the bulk action evaluated on the regular solution with boundary value $\phi_0$ and can be expressed through $\psi_\ell$ (see \textit{e.g.}~\cite{Skenderis:2002wp} for more details).

\subsection{Boundary values in terms of ambient geometry}\label{subsambound}

An old idea, which dates back to Dirac~\cite{Dirac:1935zz}, is to describe AdS and conformal fields in terms of an ambient space {$\amb$} in order to make $O(d,2)$ symmetry manifest in the sense that the group $O(d,2)$ acts linearly on {the Cartesian coordinates $X$ for} {$\amb$}. This construction also allows to have a global description of AdS and conformal spaces.
 
Let $X^A$ ($\,A=+,-,0,1,2,\ldots,d-1\,$) be the standard light-cone coordinates on $\fR^{d,2}$ so that $\eta_{+-}=1=\eta_{-+}$ and $\eta_{ab}=$ diag$(-1,+1,\ldots,+1)$ ($\,a,b=0,1,2,\ldots,d-1\,$).
We {will} often use the shortened notation $X^2=\eta_{AB}X^AX^B$.
The $AdS_{d+1}$ spacetime can be seen as the hyperboloid $X^2=-1$ of unit radius. 
In its turn, the $d$-dimensional conformal space $\manX_d$ can be identified with the projective null hypercone of light-like rays. It can also be seen as the conformal boundary of the AdS spacetime\,: $\manX_d\cong \partial ( AdS_{d+1})$. The intrinsic AdS coordinates
used in the previous subsection can be directly related to the ambient ones through $(x^a,\rho)\,=\,\big(\,X^a/X^+\,,\,(X^+)^{-2}\,\big)$ if one restricts to the domain $\rho>0$ and $X^+>0$.
The embedding of $AdS_{d+1}$ reads explicitly as 
\begin{equation}
\label{embed}
   X^a\,=\,\rho^{-\half}\,x^a\,,  \qquad X^+\,=\,\rho^{-\half}\,, \qquad X^-\,=\,-\,\half(\rho+x^ax_a)\,\rho^{-\half}\,.
\end{equation} 
In particular, the induced metric on the hyperboloid is precisely~\eqref{coord}.

The Klein-Gordon equation \eqref{ads-scalar} can be written equivalently in terms of an ambient scalar field $\Phi(X)$
satisfying
\begin{equation}
\label{adscft}
 \left(X \cdot \dl{X}+\Delta_\pm\right)\Phi_\pm=0\,, \qquad \Box\Phi_\pm=0\,.
\end{equation}
For $\Phi_\pm$ satisfying \eqref{adscft}, one can check that its value on the hyperboloid indeed satisfies~\eqref{ads-scalar} and that (at least locally) any AdS field $\varphi$ satisfying~\eqref{ads-scalar}
can be lifted by homogeneity to a $\Phi_\pm$ satisfying \eqref{adscft}.
More precisely, the correspondence between solutions to~\eqref{ads-scalar} and \eqref{adscft} is one-to-one if and only if one restricts to the domain $X^2<0$ of the ambient space.
 
The homogeneity constraint $(X \cdot \dl{X}+\Delta)\Phi=0$
defines a unique extension of a function $\overline{\Phi}(X)$ defined only on the unit hyperboloid $X^2=-1$ to a function $\Phi$ on the domain $X^2<0$, via
\begin{equation}
\label{adsamb}
 \Phi(X):=(-X^2)^{-\Delta/2}\overline{\Phi}\left(\frac{X}{\sqrt{-X^2}}\right)\,.
\end{equation}
% Geometrically, this can be easily seen via the correspondence between any point 
% $X^A$ of the domain and the point $X^{'A}=X^A/\sqrt{-X^2}$ which is the 
% intersection between the unit hyperboloid and the radial line that passes 
% through the given point. To see this explicitly, 
%let us for instance, 

The correspondence \eqref{adsamb} reads in terms of the bulk scalar field  $\varphi(x,\rho)=\rho^{\Delta/2}\phi(x,\rho)$ as follows:
\begin{eqnarray}
\label{adsamb2}
\overline{\Phi}(X^A)&=&\varphi\big(\,X^a/X^+\,,\,(X^+)^{-2}\,\big)\quad\mbox{with}\,\,X^2=-1\,,\nonumber\\ 
\Phi(X^A)&=&(X^+)^{-\Delta}\phi\big(\,X^a/X^+\,,\,-X^2(X^+)^{-2}\,\big)\,.
\end{eqnarray}
If well defined, the evaluation of the ambient field on the hypercone $X^2=0$ reads 
\begin{equation}
\label{ambdy}
\Phi(X^+x^a,X^+,-\frac12 X^+x^2)=(X^+)^{-\Delta}\phi_0(x)
\end{equation}
while its evaluation on
the hyperboloid $X^2=-1$ reads as 
%\begin{equation}
\begin{multline}
\Phi(X^+x^a,X^+,-\frac12 X^+x^2+\frac1{2X^+})~=\\=~(X^+)^{-\Delta}~\phi\big(\,\frac{X^a}{X^+},\,(X^+)^{-2}\,\big)~=~\rho^{\Delta/2}~\phi(x,\rho)\,,\,,
\end{multline} 
%\end{equation}
as can be checked by making use firstly of \eqref{adsamb2} and secondly of \eqref{embed}. 
For instance, the boundary behaviour of the AdS field is related in ambient term to the limit 
\begin{equation}
\lim_{X^{^+}\rightarrow\infty}[(X^+)^{\Delta}\Phi(X^+x^a,X^+,-\frac12 X^+x^2+\frac1{2X^+})]=\phi_0(x)\,.
\end{equation}

Another way to make contact between the ambient reformulation and the intrisic AdS treatment is to make use of the
hyperplane  $X^+=1$ in the ambient space. %Restricting to the domain $X^2<0$ 
One can map the unit hyperboloid
to the hyperplane such that the point $X^A=(X^+,X^-,X^a)$ on the hyperboloid is mapped to the point $X^{'A}=(1,X^-/X^+,X^a/X^+)$
of the hyperplane. Moreover, the coordinates 
\begin{equation}\label{coordsplane}
x^a=X^a\,,\quad \rho=-(2X^-+X^aX_a)\,,\quad 
\mbox{on the hyperplane}\,\,X^+=1\,, 
\end{equation}
are identified by this map with the intrinsic coordinates 
($\rho,x^a$) on AdS. Note that $\rho=-X^2$ on the hyperplane $X^+=1$. In particular, $\rho=0$
determines the boundary (seen as a submanifold of the hyperplane $X^+=1$).
Note also that as a coordinate on
the hyperplane $-\infty < \rho < \infty $, while as a coordinate on AdS $\rho$ can only take positive values. It can be nevertheless useful to work with fields defined also for $\rho \leq 0$. A related trick has been employed in~\cite{Vasiliev:2012vf}.

Let $\varphi(x,\rho)=\rho^{\Delta/2}\phi(x,\rho)$ be a function on AdS and $\Phi(X)$
its ambient lift satisfying $(X\cdot \d_X+\Delta)\Phi=0$ then
\begin{equation}
 \phi%(x,\rho)
\,=\,\Phi|_{X^+=1}\,, 
%\qquad \varphi(x,\rho)\,.
\end{equation} 
where the coordinates \eqref{coordsplane} are left implicit.
This gives a direct geometrical interpretation to the function $\phi$ in terms of the ambient field $\Phi$.  Moreover,
this shows why the associated boundary value $\phi_0(x)=\phi(x,0)$ 
is precisely the value of {the ambient lift} $\Phi(X)$ on the hypercone. 
Note that from this perspective it is clear why requiring $\phi(x,\rho)$ to be smooth at $\rho=0$ is natural: this is equivalent to requiring $\Phi(X)$ 
smooth at the hypercone.
Furthermore, this explains why we used the coordinates ($\rho,x^a$) rather than the Poincar\'e ones ($z,x^a$). Indeed, on the surface $X^+=1$ the expression of $z=\sqrt \rho$ in terms of $x^a,\rho$ is singular at $\rho=0$.
This singularity is on the hypercone, which is a perfectly legitimate locus in the ambient space.

In the previous paragraphs, we have implicitly assumed that the lift was performed via a homogeneity degree ``adapted'' to the asymptotic behaviour. We will now show that the 
lift of a given solution via an ``inadapted'' homogeneity degree either goes to 
zero (for the lift of subleading solutions) or to infinity on the
hypercone when $\Delta_+\neq\Delta_-$.
The ambient lift $\Phi_\pm(X):=(-X^2)^{-\Delta_\pm/2}\overline{\Phi}\left(\frac{X}{\sqrt{-X^2}}\right)$ of a given solution $\varphi$ is, in general, only well defined on the domain $X^2<0$. The lift $\Phi_\pm$ is adapted if the corresponding solution can be written as $\varphi(x,\rho)=\rho^{\Delta_{\pm}/2}\phi_\pm(x,\rho)$
where $\phi_\pm(x,\rho)$ is regular at $\rho=0$. Indeed, in such case $\Phi_\pm(X)$ is  regular on the 
entire ambient space $\fR^{d,2}\backslash \{ 0 \}$. However, the ``inadapted'' lift $\Psi_\mp:=(-X^2)^{-\Delta_\mp/2}\overline{\Phi}\left(\frac{X}{\sqrt{-X^2}}\right)$ of the same AdS field $\varphi$ can be written as $\Psi_\mp(X)=(X^2)^{(\Delta_\pm-\Delta_\mp)/2}\Phi_\pm(X)$. The evalutation of the right-hand-side at $X^2=0$ either vanishes or blows up (when $\Delta_+\neq\Delta_-$) since the factor $\Phi_\pm(X)$ is regular on the hypercone.

The particular case of the critical scalar field corresponds to the equations \eqref{adscft} for $\Delta_\pm=\frac{d}2\pm \ell$.
For $\Delta_-=\frac{d}2-\ell$, the space of solutions of \eqref{adscft}, \textit{i.e.} $\left(X \cdot \dl{X}+\Delta_-\right)\Phi_-=0$ and $\Box\Phi_-=0$,
can be quotiented by the equivalence relation
\begin{equation}\label{3cstrs}
\Phi_-\,\sim\,\Phi_-\,+\,(X^2)^\ell\,\Psi_+\,,
\end{equation}
where the ambient field $\Psi_+(X)$ is subject to
the conditions $\left(X \cdot \dl{X}+\Delta_+\right)\Psi_+=0$ and $\Box\Psi_+=0$ and
is related to the lift of the subleading solutions.
Therefore the two equations \eqref{adscft} together with the {equivalence} \eqref{3cstrs} provide an ambient description of the higher-order singleton. 
Evaluating the fields $\Phi_-(X)$ on the hyperboloid $X^2=-1$ gives its realization as an on-shell bulk scalar field $\varphi$. 
Another option is to evaluate $\Phi_-(X)$ on the hypercone $X^2=0$ to obtain its realization as an on-shell boundary scalar field $\phi_0$. We will discuss in more details the ambient descriptions of the (higher-order) singleton in Subsection \bref{higher-ordersinglet}.

Although the present discussion is restricted to the case of a scalar field these aspects of the boundary behavior
are nearly unchanged if one considers gauge fields on AdS. The only difference is that in the case of gauge fields
the conformal dimensions $\Delta_{\pm}$ are restricted by gauge invariance to integer values. 
What needs to be carefully reconsidered in the case of gauge fields is the precise definition of asymptotic behavior. 
Possible subtleties have to do with the tensorial nature of the involved gauge fields and parameters. 
One must prescribe boundary behavior not only for gauge fields but also for gauge parameters and (higher) reducibilty relations (if any). 
It turns out that these subtelties can be resolved through using the ambient approach in combination with the BRST technique. 
Indeed, in the ambient space there is a preferable coordinate system (the Cartesian coordiantes where $\mathfrak{o}(d,2)$ acts linearly) 
so that there is a simple way to define asymptotic behaviour of tensor fields and moreover the conformal boundary identifies simply with (strictly speaking, the quotient of) the submanifold of the ambient space rather than the asymptotic boundary. Furthermore, fixing the asymptotic behavior in BRST terms automatically does so for gauge fields, paremeters, \textit{etc}, 
in a way consistent with gauge invariance.

\section{Group theoretical preliminaries}\label{preliminaries}

The group-theoretical catalogue, given in Subsection \bref{catalog},
of the $\mathfrak{o}(d+2)$-modules that are relevant for the present paper
may help to put some order and fix the terminology in the following discussions.
When possible, it has been checked to fit within the known classification
of generalized Verma $\mathfrak{o}(d+2)$-modules \cite{Shaynkman:2004vu} (nevertheless,
the details of these explicit checks is omitted here for the sake of brevity because they are technical
and correspond to standard results).

\subsection{Generalized Verma $\mathfrak{o}(d,2)$-modules}\label{Verma}

In the following, we will not be much concerned about unitarity issues therefore one can work with complex algebras and the choice of signature is somewhat irrelevant for the classification of generalized Verma $\mathfrak{o}(d+2)$-modules.
However, for definiteness one chose the two-time signature $\mathfrak{o}(d,2)$ relevant for AdS/CFT discussions.
The Lie algebra $\mathfrak{o}(d,2)=\mbox{span}
\{\textsc{J}_{AB}\}$ can be presented by its {generators} $\textsc{J}_{AB}=-\textsc{J}_{BA}$ (where $A,B=0,0^\prime,1,2,\ldots,d$) modulo the {commutation relations} $\left[\textsc{J}_{AB},\textsc{J}_{CD}\right]\,=\,i\,\eta_{BC}\textsc{J}_{AD}\, +\,$ antisymmetrizations, with
$\eta_{AB}\,=\,\,$diag$(-1,-1,+1,+1,\ldots,+1)$.

The maximal compact subalgebra of $\mathfrak{o}(d,2)$ is the direct sum
$\mathfrak{o}(2)\oplus\mathfrak{o}(d)$.
In the $AdS_{d+1}$ spacetime endowed with the usual global coordinates (or on the conformal boundary $\partial AdS_{d+1}$ whose topology is roughly $S^1\times S^{d-1}$), the summand $\mathfrak{o}(2)$ corresponds to time translations generated by the (conformal) Hamiltonian $\textsc{E}=\textsc{J}_{0' 0}$
while $\mathfrak{o}(d)$ corresponds to the rotations generated by $\textsc{J}_{ij}$ (where $i,j=1,2,\ldots,d-1$).
The remaining generators can be recast in the form of {ladder operators} $\textsc{J}^{\pm}_j=\textsc{J}_{0j}\mp i \textsc{J}_{0'j}\,,$ raising or lowering the (conformal) {energy} (= eigenvalue of $\textsc{E}$) by one unit, due to the commutation relations
\begin{eqnarray}
&&\left[\textsc{E}, \textsc{J}_i^{\pm}\right]=\pm  \textsc{J}_i^{\pm} ~~~;~~~ \left[\textsc{J}_{ij},\textsc{J}^{\pm}_k\right]= 2i\delta_{k[j} \textsc{J}^{\pm}_{i]}\nonumber\\
&&\left[\textsc{J}^-_i, \textsc{J}^+_j\right]=2(i\textsc{J}_{ij}+\delta_{ij}\textsc{E})\label{comrel}\\
&&\left[\textsc{J}_{ij},\textsc{J}_{kl}\right]=i\delta_{jk}\textsc{J}_{il} \,+\, \mbox{antisymetrizations}\nonumber
\end{eqnarray}

Following standard usage in the litterature (although the compact subalgebra is often priviledged), we will often make use of the CFT language to describe the $\mathfrak{o}(d,2)$-modules.
Since the signature is not directly relevant for most of our considerations,
all results can be directly translated in CFT language either by making use of the radial quantization or
by priviledging the
$\mathfrak{o}(1,1)\oplus\mathfrak{o}(d-1,1)$ subalgebra where
$\mathfrak{o}(1,1)$ correspond to scale transformations
generated by $\textsc{D}:= \textsc{J}_{0' d}$ and
$\mathfrak{o}(d-1,1)$ to the Lorentz transformations generated
by $\textsc{J}_{ab}$ (where $a,b=0,1,2,\ldots,d-1$)
of the conformal $d$-dimensional space.
The ladder operators correspond to the translation generators
$\textsc{P}_a=\textsc{J}_{d\,a}+\textsc{J}_{0'a}$ and the conformal boost generators $\textsc{K}_a=\textsc{J}_{d\,a}- \textsc{J}_{0'a}$ respectively raising and lowering the conformal weight (= eigenvalue of $\textsc{D}$) by one unit, due to the analogue of the commutation relations \eqref{comrel}.

\vspace{3mm}

\noindent\textbf{Notations:}

\vspace{2mm}

\noindent $\bullet$ The finite-dimensional irreducible $\mathfrak{o}(2)\oplus\mathfrak{o}(d)$-module characterized by the ``energy'' $\Delta$ for $\mathfrak{o}(2)$
(or ``conformal weight'' in CFT language) and the ``spin'' represented by a Young diagram $Y$ for $\mathfrak{o}(d)$,
 will be denoted as
${\cal Y}\left(\Delta,Y\right)$. A Young diagram $Y$ made of a single row of length $r$ will be denoted as the number $Y=r$ for simplicity.

\vspace{2mm}

\noindent $\bullet$ The (generalized) Verma $\mathfrak{o}(d,2)$-module ${\cal V}\left(\Delta,Y\right)$ is defined as follows:
\begin{equation}\label{Vmoddef}
{\cal V}\left(\Delta,Y\right)={\cal U}\Big({\mathbb R}^{d}\Big)\tensor {\cal Y}\left(\Delta,Y\right)\,,\quad {\mathbb R}^{d}=\mbox{span}\{\textsc{J}^+_i\}\,,
\end{equation}
where ${\cal U}\Big({\mathbb R}^{d}\Big)$ stands for the enveloping of the abelian subagebra generated by $\textsc{J}^+_i$. In other words,
the module is generated by the free action of the raising operators $\textsc{J}^+_i$ on the submodule ${\cal Y}\left(\Delta,Y\right)$, which is a finite-dimensional irreducible $\mathfrak{o}(2)\oplus\mathfrak{o}(d)$-module (the conformal primary in CFT language).
As one can see, $\Delta$ is the lowest energy in the module $\mathfrak{o}(2)$-decomposition.

\vspace{2mm}

\noindent $\bullet$ Following the standard notation in the physics community, the $\mathfrak{o}(d,2)$-irreducible quotient
of the Verma module ${\cal V}\left(\Delta,Y\right)$
will be denoted by ${\cal D}\left(\Delta,Y\right)$.

\subsection{Group-theoretical catalogue of $\mathfrak{o}(d,2)$-modules}\label{catalog}

The Verma $\mathfrak{o}(d,2)$-modules are classified according to their lowest weights ($\Delta,Y$). Some of the relevant
$\mathfrak{o}(d,2)$-modules reviewed below are, strictly speaking, not Verma modules. Nevertheless, the modules have been
sorted by the labels ($\Delta,Y$). The terminology adopted aims to make contact with their standard field-theoretic realization
in the litterature, either as conformal or AdS fields.
In the description of the realization in terms of conformal fields, the $\mathfrak{o}(d,2)$-modules will be described by the corresponding conformal primary field but of course the whole module is spanned by the primary together with all its (nontrivial) descendants.

\vspace{5mm}

Let $\ell\in{\mathbb N}-\{0\}$ and $s,t\in{\mathbb N}$.

\vspace{5mm}

\noindent $\bullet$
%\subsubsection*
\underline{$(\Delta_\pm, Y)=(\frac{d}2\pm \ell\,,\, 0)$\,:}

\vspace{3mm}

\textbf{Higher-order scalar singleton:}
The Verma module ${\cal V}\left(\frac{d}2-\ell,0\right)$ is reducible though indecomposable, it is isomorphic to the gluing
\begin{equation}
{\cal V}\left(\frac{d}2-\ell,0\right)\cong{\cal D}\left(\frac{d}2-\ell,0\right)
\niplus
 {\cal V}\left(\frac{d}2+\ell,0\right)\,,
\end{equation} 
where the quotient
\begin{equation} \label{highderscalquot}
{\cal D}\left(\frac{d}2-\ell,0\right)\cong \frac{{\cal V}\left(\frac{d}2-\ell,0\right)}{{\cal V}\left(\frac{d}2+\ell,0\right)}
\end{equation}
is irreducible.
The module ${\cal V}\left(\frac{d}2-\ell,0\right)$ corresponds to the off-shell scalar field $\phi(x)$ on the flat conformal space $\manX_d$
with conformal weight $\Delta_-=\frac{d}2-\ell$. Its submodule ${\cal V}\left(\frac{d}2+\ell,0\right) \subset {\cal V}\left(\frac{d}2-\ell,0\right)$ corresponds to the descendant $\Box_0^\ell\phi(x)$. Thus the irreducible quotient
${\cal D}\left(\frac{d}2-\ell,0\right)$ corresponds to an on-shell conformal scalar field subject to the polywave equation {$\Box_0^\ell\phi=0$}
of order $2\ell$. The module ${\cal D}\left(\frac{d}2-\ell,0\right)$ will be called the \textit{scalar singleton of order $\ell$}, since for $\ell=1$ it is the usual scalar singleton (also called ``Rac'') subject to the usual wave equation of second order. In the terminology of \cite{Iazeolla:2008ix}, they would be called $\ell$-linetons because their weight decomposition is described by $\ell$ lines in the weight-space diagram.

\textbf{Critical AdS scalar:} Consider the Klein-Gordon equation $\left(\nabla^2-m^2\right)\varphi=0$
for a scalar field on $AdS_{d+1}$ with $\nabla^2$ the Laplace-Beltrami operator of $AdS_{d+1}$.
The space of solutions of the Klein-Gordon equation
on $AdS_{d+1}$ with critical mass-square $m^2=\ell^2-(\frac{d}2)^2$ is the semidirect sum
\begin{equation}
{\cal V}\left(\frac{d}2-\ell,0\right)\niplus
\,{\cal V}\left(\frac{d}2+\ell,0\right)
\end{equation} 
where the above two modules of solutions correspond to the two possible boundary behaviours $\varphi_\pm\sim \rho^{d\pm 2\ell}\phi_\pm$ for $\rho\sim 0$.
If one considers the quotient of this Klein-Gordon solution space by the solutions with subleading boundary behaviour $\varphi_+\sim \rho^{d+2\ell}\phi_+$, then even the leading boundary data with nontrivial $\Box^\ell\phi_-$ can be gauged away. The quotient space
is actually isomorphic to the irreducible module
${\cal D}\left(\frac{d}2-\ell,0\right)$ and describes the singleton of order $\ell$ as a critical AdS scalar field whose bulk degrees of freedom can be gauged away by the above quotient (see \cite{Flato:1980we} for the seminal paper {devoted} to the case of $d=3$ and $\ell=1$).

\vspace{5mm}

\noindent $\bullet$ 
%\subsubsection*
\underline{$(\Delta%_+
,Y)=(d+s-t-1,s)$ %with $0< t\leqslant s$
\,:}
\vspace{3mm}

\textbf{Partially conserved current:} The Verma module ${\cal V}\left(d+s-t-1,s\right)$ corresponds to a rank-$s$ symmetric traceless tensor fields
$j^{a_{1}\ldots \,a_{s}}$ with conformal weight $\Delta_+=d+s-t-1$.
For $t>s$, this module is irreducible so that we denote it as
${\cal D}\left(d+s-t-1,s\right)={\cal V}\left(d+s-t-1,s\right)$.
However, for $0<t\leqslant s$, the Verma module %${\cal V}\left(d+s-t-1,s\right)$
is
reducible, though indecomposable, because it contains the submodule 
\begin{equation}
{\cal I}\left(d+s-1,s-t\right)\subset{\cal V}\left(d+s-t-1,s\right)
\end{equation} 
isomorphic to the Verma module ${\cal V}\left(d+s-1,s-t\right)\cong {\cal I}\left(d+s-1,s-t\right)$
which corresponds to the partial conservation law of depth $t$ (\textit{i.e.} {$t$-th} divergence of the current:
$\partial_{a_1} \cdots\partial_{a_t}j^{a_{1}\ldots \,a_{s}}$), but one can consider the quotient
\begin{equation}\label{partconscurrmod}
{\cal D}\left(d+s-t-1,s\right)\cong \frac{{\cal V}\left(d+s-t-1,s\right)}{{\cal V}\left(d+s-1,s-t\right)}
\end{equation}
The irreducible module ${\cal D}\left(d+s-t-1,s\right)$
will be called a partially conserved current of spin $s$ and depth $t$
(discussed \textit{e.g.} in \cite{Dolan:2001ih})
since they are the generalisation of the conformal conserved currents (case $t=1$).
More concretely, a \textit{spin-$s$ and depth-$t$ partially conserved current} is a primary field
$j^{a_1\ldots a_s}(x)$ on $\manX_d$ with weight $\Delta=d+s-t-1$, which is
symmetric, traceless and partially conserved:
\begin{equation}
\label{current}
 \left(\dl{p}\cdot\dl{p}\right)j(x,p)=0\,, \qquad \left(\dl{x}\cdot\dl{p}\right)^tj(x,p)=0\,,
\end{equation}
{where by making use of an auxiliary variable $p^a$ the tensor $j_{a_1\ldots a_s}(x)$ has been packed into a generating function $j(x,p)=j_{a_1\ldots a_s}(x)p^{a_1}\ldots p^{a_s}$.}

\vspace{5mm}

\noindent $\bullet$ 
%\subsubsection*
\underline{$(\Delta%_-
,Y)=(1+t-s,s)$% with $0< t\leqslant s$
\,:}

\vspace{3mm}

\textbf{Higher-depth shadow field:}
A \textit{spin-$s$ and depth-$t$ shadow field} is a primary field $\phi_{a_1\ldots a_s}(x)$ on $\manX_d$ with weight $t+1-s$, symmetric, traceless and subject to generalized Fradkin--Tseytlin (FT) gauge transformations:
\begin{equation}
\label{shadow}
 \left(\dl{p}\cdot\dl{p}\right)\phi(x,p)=0\,, \qquad \delta_\varepsilon \phi(x,p)=\Pi \left(\,(p\cdot\dl{x})^t\varepsilon(x,p)\,\right)\,,
\end{equation}
where $\Pi$ denotes the projection to the traceless component and the tensor has been packed into a generating function.
Since the shadow field of spin $s$ and depth $t$ does not correspond to a genuine Verma module
(but somehow to the contragradient of a Verma module), it will be denoted as
${\cal W}\left(1+t-s,s\right)$ where the primary field is a rank $s$ traceless symmetric tensor field with gauge symmetries which are generalized FT transformations
(projected $t$th gradient of a rank $s-t$ traceless symmetric tensor field).

\textbf{Higher depth Fradkin--Tseytlin field:} For $d$ even, there is a submodule isomorphic to ${\cal D}\left(d+s-t-1,s\right)$ of
the indecomposable module ${\cal W}\left(1+t-s,s\right)$.
This submodule is generated by the descendant corresponding to the higher depth version of FT equations of the form
\begin{equation}\label{genFT}
\Box^\frac{d-4}2\partial_{b_1} \cdots\partial_{b_{s-t+1}}{\cal W}^{a_{1}\ldots \,a_{s}|b_{1}\ldots \,b_{s-t+1}}+\ldots\,=\,0
\end{equation}
where ${\cal W}^{a_{1}\ldots \,a_{s}|b_{1}\ldots \,b_{s-t+1}}$
is the generalized Weyl tensor of the shadow field $\phi_{a_1\ldots a_s}(x)$ (\textit{i.e.} it is the
traceless part of its generalized Riemann tensor, which is in turn the ($s-t+1$)-th curl of the shadow field) and the dots stand for extra terms involving the same number of derivatives of the generalized Weyl tensor 
but with more complicated contractions of indices. So one can consider the quotient
\begin{equation}
{\cal C}\left(1+t-s,s\right)= \frac{{\cal W}\left(1+t-s,s\right)}{{\cal D}\left(d+s-t-1,s\right)}
\end{equation} 
which will be called a \textit{spin-$s$ and depth-$t$ Fradkin--Tseytlin field}
and which can be seen as an on-shell higher-depth shadow field. These equations fit into the classification~\cite{Shaynkman:2004vu} of conformal equations. They also belongs to the class of conformal gauge fields
considered in~\cite{Vasiliev:2009ck} where the gauge invariant Lagrangians in terms of Weyl tensors have been constructed.

\vspace{2mm}

\textbf{Partially massless AdS field:}
In the metric-like formulation \cite{Deser:2001pe}, the free \textit{spin-$s$ depth-$t$ partially massless fields}
on $AdS_{d+1}$ are described via rank-$s$ symmetric tensor gauge fields
$\varphi_{\mu_1\ldots \mu_s}$ modded by gauge transformations of the form
\begin{equation}
\delta_\varepsilon \varphi_{\mu_1\ldots \mu_s}=\nabla_{(\mu_1}\ldots \nabla_{\mu_t}\varepsilon_{\mu_{t+1}\ldots \mu_{s)}} + \mbox{terms involving the AdS metric},
\end{equation}
and subject to second order field equations generalizing Fronsdal ones.
Indeed, when the depth $t$ is equal to one, these fields are usually called massless
and were described  by Fronsdal \cite{Fronsdal:1978vb}.
The space of inequivalent solutions of PM field equations is the direct sum
$${\cal D}\left(d+s-t-1,s\right)\oplus\,{\cal W}\left(1+t-s,s\right)$$
where the above two modules of solutions correspond to the two possible boundary behaviours $\varphi_\pm\sim \rho^{\Delta_\pm/2}\phi_\pm$
for $z\sim 0$. The partially conserved currents appear as boundary values of ``normalizable'' solutions of these equations,
while the higher-depth shadow fields appear as
boundary values of (so-called) ``non-normalizable'' solutions of the same equations (see \textit{e.g.} \cite{Metsaev:1999ui} for the massless case $t=1$).
The \textit{global reducibility parameters of partially massless fields} are, by definition, gauge parameters $\varepsilon_{\nu_1\ldots \nu_{s-t}}$ such that the corresponding gauge transformation vanishes,
$\delta_\varepsilon \varphi_{\mu_1\ldots \mu_s}=0$. Such global reducibility parameters are analogous but distinct from the generalized Killing tensors introduced in
\cite{Nikitin} because in the latter generalized Killing equation no term involving the metric appears.

\vspace{5mm}

\noindent $\bullet$ 
%\subsubsection*
\underline{$(\Delta%_-
,Y)=(1-s, s-t)$\,:}
\label{sec:group-Killing}
\vspace{3mm}

\textbf{Higher-depth conformal Killing tensors:} For $s\geqslant t$, a \textit{rank-($s-t$) and depth-$t$ conformal Killing tensor}\footnote{They would be called generalized conformal Killing tensor fields of rank $r=s-t$ and of order $t$ in the terminology of \cite{Nikitin}.} is a symmetric traceless tensor field $\varepsilon_{b_1\ldots \,b_{s-t}}(x)$ on
$\manX_d$
solution of the generalized conformal Killing equation \cite{Nikitin}
\begin{equation}\label{genKeq}
 \partial_{(a_1} \cdots\partial_{a_t}\epsilon_{a_{t+1}\ldots \,a_s)}(x)=g_{(a_1a_2}(x)\chi_{a_3\ldots \,a_s)}(x)\,.
\end{equation}
The (usual) conformal Killing tensors are the particular case of depth one.
The (higher-depth) Killing tensor fields on flat spacetime are pertinent for conformal HS gravity because they are the \textit{global reducibility parameters of (higher-depth) shadow fields} linearized around the flat background.
The space of the rank-($s-t$) and depth-$t$ conformal Killing tensors  is isomorphic to the
finite-dimensional irreducible $\mathfrak{o}(d)\oplus \mathfrak{o}(2)$-module
${\cal Y}\big(\,1-s\,,\,(s-1,s-t)\,\big)$ labeled by a Young diagram $Y=(s-1,s-t)$ with first row of length $s-1$
and second row of length $s-t$.
This module is also isomorphic to the irreducible $\mathfrak{o}(d,2)$-module, obtained as the
 quotient
$${\cal D}\left(1-s,s-t\right)\cong \frac{{\cal V}\left(1-s,s-t\right)}{{\cal I}\left(1+t-s,s\right)}\cong {\cal Y}\Big(\,1-s\,,\,(s-1,s-t)\,\Big)$$
where the Verma module
${\cal V}\left(1-s,s-t\right)$ corresponds to the gauge parameters $\varepsilon_{b_1\ldots \,b_{s-t}}(x)$
in the FT transformations \eqref{shadow}
for a shadow field $\varepsilon_{a_1\ldots \,a_s}(x)$ of spin $s$ and depth $t$,
while the submodule ${\cal I}\left(1+t-s,s\right)$
corresponds to the higher-depth conformal Killing equation \eqref{genKeq}
equivalent to the reducibility condition $\Pi \left(\,(p\cdot\dl{x})^t\varepsilon(x,p)\,\right)=0$.
The module of (higher-depth) conformal Killing tensors is actually also isomorphic to
the space of global reducibility parameters of (partially) massless fields.

\subsection{Generalized Flato--Fronsdal theorem}

\begin{prop}
\label{prop:FF}
The tensor product of two singletons of order $\ell$($\geqslant 1$) decomposes as the following sum
\begin{equation}
{\cal D}\Big(\ffrac{d}{2}-\ell,0\Big)\otimes\,{\cal D}\Big(\ffrac{d}{2}-\ell,0\Big)
\,=\,\bigoplus\limits_{s=0}^\infty\bigoplus\limits_{k=1}^{\ell}{\cal D}(d+s-2k,s)
\end{equation} 
of the irreducible $\mathfrak{o}(d,2)$-modules describing the partially conserved currents of all ranks $s\in\mathbb N$ and all odd depths $t$ (= $2k-1$)
ranging from $1$ to $2\ell-1$.
\end{prop}

For $\ell=1$, this is the Flato--Fronsdal theorem for the tensor products of two scalar singletons \cite{Flato:1978qz,Vasiliev:2004cm}.~\footnote{We are gratefull to E.~Skvortsov for informing us that Proposition~\bref{prop:FF} has been also independently obtained by him and collaborators.} The proof is relegated to the Appenix~\bref{sec:FFproof}.

\section{Partially massless fields}\label{Partiallymassless}

Though the phenomenon of ``partial masslessness'' on (A)dS was first observed for fields of spin $2$ by Deser and Nepomechie \cite{Deser:1983tm,Deser:1983mm}, the term was coined later %by Deser and Waldron
when the higher-spin generalization was sketched \cite{Deser:2001pe} (the harmonic analyses of Higuchi \cite{%Higuchi:1986py,
Higuchi:1986wu%,Higuchi:1989gz
} which was obtained in the meanwhile provides a useful mathematical complement).
Over the years, various formulations of free PM fields have been investigated, such as metric-like
\cite{Zinoviev:2001dt},  frame-like \cite{Skvortsov:2006at,Zinoviev:2008ze} and BRST-parent \cite{Alkalaev:2009vm,Grigoriev:2011gp,Alkalaev:2011zv} approaches.
More recently, the possibity of consistent interactions has been studied 
at the level of cubic vertices for any integer spins \cite{Joung:2012rv,Joung:2012hz,Boulanger:2012dx}.
An AdS/CFT dictionary for PM fields was sketched in \cite{Dolan:2001ih} (see \cite{Deser:2003gw} for a dS/CFT analogue).

\subsection{Ambient representation of partially massless fields}

Consider symmetric tensor fields $\Phi_{A_1\ldots A_s}(X)$ defined on the ambient space $\fR^{d,2}\backslash\{0\}$.
Identify them as Taylor coefficients in the power series expansion of a generating function
$\Phi(X,P)\,=\,\sum_s\,\frac1{s!}\, \Phi_{A_1\ldots A_s}(X)\,P^{A_1}\ldots P^{A_s}$ where the $P$'s are mere auxiliary variables. The homogeneity degree in $P$ corresponds to the rank of the tensor field. In addition to the action of $\mathfrak{o}(d,2)$ by
$J_{AB}=X_{A} \dl{X^{B}}-X_{B} \dl{X^{A}}+P_{A} \dl{P^{B}}-P_{B} \dl{P^{A}}$ the space of such fields is equipped with an action 
of the algebra $\mathfrak{sp}(4)$ generated by the 10 operators
\begin{equation}
 \begin{gathered}
\Box=\half \d_{X}\cdot\d_{X}\,, \qquad    S=\d_X\cdot\d_P\,,\qquad T=\half \d_{P}\cdot\d_{P}\,,
\\
\bsd=X\cdot\d_{P}\,,\qquad  S^\dagger=P\cdot\d_{X}  \\
U_-=P\cdot\d_{P}-X\cdot\d_{X}\,,\qquad U_+=P\cdot\d_{P}+X\cdot\d_{X}+d+2\,,\\
\bar{\Box}=\half X^2,\qquad \bs=X\cdot P\,,\qquad \bar T=\half P\cdot P\,.
\end{gathered}
\label{sp4}
\end{equation}
There are two obvious automorphisms induced by $P\to -\dl{P}$, $\dl{P}\to P$ or $X\to -\dl{X}$, $\dl{X}\to X$ which  are somehow the analogue of Fourier transformation, respectively in the space of auxiliary variable $P$
or in the space of ambient coordinates $X$.
The two operators counting the homogeneity degree in $X$ and $P$ respectively, $N_X=X\cdot\d_{X}$ and $N_P=P\cdot\d_{P}$ will sometimes be used below, although they
are linear combinations of $U_\pm$.

In ambient terms, the totally symmetric depth-$t$ partially massless fields
can be described through the following constraints and gauge symmetries~\cite{Alkalaev:2009vm,Alkalaev:2011zv}:
\begin{equation}
\label{pmassl-1}
\begin{gathered}
 T\Phi=S\Phi=\Box\Phi=(U_--(t+1))\Phi=0\,,\quad  (\bsd)^t\Phi=0\,, \\
 \quad \Phi\simeq \Phi+\sd \chi \,, 
\end{gathered}
\end{equation}
where $\chi$ satisfies the same constraints but with $t+1$ replaced with $t-1$. To describe a spin $s$ field one in addition imposes $(P\cdot \d_P-s)\Phi=0$ and $(P\cdot \d_P-s+1)\chi=0$. To check the consistency it is useful to employ the relations of $\mathfrak{sp}(2)$ generators $E,F,H$ in the standard basis $\commut{H}{E}=2E$, $\commut{H}{F}=-2F$, $\commut{E}{F}=H$
\begin{equation}
\label{sp2}
 \commut{E}{F^t}=F^{t-1}t(H-t+1)\,, \qquad \commut{E^t}{F}=E^{t-1}t(H+t-1)\,,
\end{equation}
and observe that the 3 operators $\sd,\bsd,U_-$ satisfy the same algebra.

It can be also convenient to work with the partially gauge fixed version of the same system~\cite{Joung:2012rv}
\begin{equation}
\label{pmassl}
\begin{gathered}
 T\Phi=S\Phi=\Box\Phi=(U_--(t+1))\Phi=0\,,\quad  \bsd\Phi=0\,, \\ \Phi\simeq \Phi+(\sd)^t \chi\,,
 \end{gathered}
\end{equation}
where $\chi$ satisfies the same constraints but with $t+1$ replaced with $1-t$.  
Here, the consistency is easy to check by making use of the second relation of~\eqref{sp2}. 
Note that \eqref{pmassl} can be seen as a 
straightforward ambient rewriting of the PM equations of motion and gauge 
transformations in terms of intrinsic AdS coordinates. 

To obtain this description from~\eqref{pmassl-1}, one first uses the partial gauge 
$\bsd \Phi=0$. The parameter preserving the gauge condition satisfies 
$\bsd \sd \chi=0$ which together with $U_-\sd\chi=(t+1)\sd\chi$ implies 
$\sd\chi=(\sd)^t\lambda$ for some $\lambda$ satisfying $\bsd\lambda=0$ and 
$U_-\lambda =(1-t)\lambda$ along with $\Box \lambda=T\lambda=S\lambda=0$. Let us spell out the argument explicitly in the simplest case $t=1$.
Using $U_-\chi=\chi$ the gauge variation of the condition $\bsd \Phi=0$ takes the form
$\delta (\bsd \Phi)=\bsd \sd \chi= -\chi +\sd\bsd \chi$.  It follows the gauge is reachable and moreover 
parameters preserving the gauge can be represented as $\chi=\sd\lambda$ with $\bsd\lambda=0$.

In~\eqref{pmassl-1} and \eqref{pmassl}
the homogeneity degree in $X$ is set to $s-t-1=-\Delta_-$ (for a spin-$s$ depth-$t$ field). An alternative description based on $\Delta_+=d+s-t-1$ is also possible and will be discussed in Section~\bref{sec:current}. Just like in the case of a massive scalar both descriptions are equivalent in the domain $X^2<0$  while in the vicinity of $X^2=0$ they correspond to boundary values with the respective asymptotics.

In these formulations the gauge parameters are subject to differential
constraints. There is an elegant way to replace the system with the genuine
gauge one where no differential constraints are imposed on gauge parameters. The
idea is to implement all the constraints through the BRST operator. The simple
rule is to represent ghost variables associated to constraints in coordinate
representation while those associated to gauge generators in momentum representation so that
the physical fields are found at degree zero.

The BRST operator implementing~\eqref{pmassl-1} reads as
\begin{equation}
\label{PM-BRST}
 \Omega=c_0\Box+c S+ \xi T+\sd \dl{b}+\mu (U_--t-1+2b\dl{b})+\nu (\bsd)^t+G\,, \quad
\end{equation}
where the ghost variables $c_0,c,\xi,\mu,\nu$ of ghost degree $1$ and $b$ with $\gh{b}=-1$ have been introduced. 
Here, $G$ denotes the terms encoding higher structures of the gauge algebra
and this operator is chosen  such that $G1=0$. In particular, {this condition
determines the constant term and the ghost contribution of the constraint that enters $\Omega$ as the term linear in $\mu$.} Note that the ghost variables associated to $\sd$ are represented in momentum representation to reflect that $\sd$ generates a gauge transformation (see \textit{e.g.}~\cite{Grigoriev:2011gp,Bekaert:2012vt} for more details on the BRST implementation of constraints).

The equations of motion and gauge symmetries are
\begin{equation}
 \Omega \Psi=0\,, \qquad \Psi \simeq \Psi+\Omega \Lambda\,, \qquad \gh{\Psi}=0\,,\quad \gh{\Lambda}=-1
\end{equation}
Note that the explicit expression for $\Psi$ and $\chi$ reads as
\begin{equation}
 \Psi=\Phi+c_0b\, \Phi_\Box+cb\, \Phi_S+\xi b\,\Phi_T+\mu b\,\Phi_U+\nu b \,\Phi_{\bsd}\,,\qquad \Lambda=b \lambda
\end{equation}
where all the component fields depend on $X^A,P^A$. 
We see that using BRST procedure produces specific set of auxiliary fields. 
Moreover,  $\lambda$ is not subject to any constraints and hence the above system is a genuine gauge system.

To complete the ambient description for PM fields one shoudl also specify the domain where $\Psi$ and $\Lambda$ are defined.
We take it to be the neigbourhood of the hyperboloid (this in turn is equivalent to taking $\Psi$ defined on the domain $X^2<0$, see~\cite{Bekaert:2012vt} for more details).

\subsection{Parent formulation of partially massless fields}

An alternative way to replace the partially gauge fixed formulation~\eqref{pmassl-1} (or \eqref{pmassl})
with a genuine gauge invariant one was proposed in~\cite{Alkalaev:2009vm,Alkalaev:2011zv} (see also~\cite{Barnich:2006pc}) and is based on the so-called parent formulation.

The idea is to introduce extra fields by, roughly speaking, putting the ambient space system~\eqref{pmassl-1} to the target space. More precisely, replacing $X^A$ with a formal variable $Y^A$ and allowing the generating function $\Psi$ to depend on local AdS coordinates $x^\mu$ and their differentials $\theta^\mu \equiv dx^\mu$. The Grassmann variables $\theta^\mu$ are to be understood as ghost variables with $\gh{\theta^\mu}=1$.

The parent system is determined by
\begin{equation}
\label{constr-inter}
 \begin{gathered}
 \Omega=\nabla+Q\,, \qquad \Psi=\Psi(x,\theta|Y,P,b)\,, \\
\Box\Psi=S\Psi=T\Psi=(\bsd\Psi)^t=(U_--t-1+2b\dl{b})\Psi=0\,,                                                                                                                               \end{gathered}
\end{equation}
where in the constraints of the second line and BRST operator $Q=\sd\dl{b}$
implementing the gauge equivalence from~\eqref{pmassl-1}
the replacement $X^A \to Y^A+V^A$ and $\dl{X^A}\to \dl{Y^A}$ has been made and
$\nabla$ is given by
\begin{multline}
%\label{nabla}
 \nabla=\derham-e^A\dl{Y^A}- \omega^B_A\Big(Y^A\dl{Y^B}+P^A\dl{P^B}\Big)
=\\
=\derham-\derham V^A\dl{Y^A}-
\omega^B_A\Big(\big(\,Y^A+V^A\,\big)\dl{Y^B}+P^A\dl{P^B}\Big)\,.
 \label{covder}
\end{multline}
Here, $\derham=\theta^\mu \dl{x^\mu}$ is the De Rham differential, $\omega^A_B=\theta^\mu \omega_{\mu B}^A$ is a  flat connection taking values in $\mathfrak{o}(d,2)$ and $V^A$ a fixed section,
called compensator, and $e^A=\nabla V^A$ is a frame field. It is assumed that $e_\mu^A=\nabla_\mu V^A$ has a maximal rank. To describe AdS fields one takes $x^\mu$ to be local coordinates
on $AdS_{d+1}$, $\omega^B_A$ to be an $\mathfrak{o}(d,2)$-connection on AdS, and the compensator $V^A$ to be such that $V^2=-1$.  
Note that the spacetime derivatives $\dl{x_\mu}$ enters the operator $\Omega$ only through $\nabla$ so that all the equations of motion, gauge generators, etc, of the former are manifestly first order in spacetime derivatives.

Introducing ghost degree zero component fields according to $\Psi=F+b \theta^\mu A_\mu$ the equations of motion and gauge symmetries read explicitly as
\begin{equation}
 \nabla A=0\,, \qquad \nabla F=\sd A\,, \qquad \delta A=\nabla \lambda \,,\quad \delta F=\sd \lambda\,,
\end{equation}
along with the target space constraints from the second line of~\eqref{constr-inter}. In this form it is easy to relate the formulation to the partially gauge fixed ambient space system~\eqref{pmassl-1}. Indeed,
at least locally the first equation can be solved as $A=\nabla \chi$ so that $A$ can be gauged away. In this gauge $A=0$ and the second equation and the residual gauge invariance take the form
\begin{gather}
\label{parent-gf}
 \nabla F=0\,,
 \quad \Box F=S F=T F=(\bsd)^t F=(U_--t-1)F=0\,,\\
 \label{parent-gf-gs}
  \delta F=\sd \lambda\,,\quad  \nabla \lambda=\Box \lambda=S \lambda=T \lambda=(\bsd)^t \lambda=(U_--t+1)\lambda=0
\end{gather}
This system can be seen as a pulback of a system defined on the ambient space $\amb$ to the AdS hyperboloid. In particular,
$\nabla$ is a pullback of the standard flat connection $\nabla_0$ on $\amb$. In a suitable frame $\nabla_0=\theta^A(\dl{X^A}-\dl{Y^A})$ and the above system reduces to~\eqref{pmassl-1} upon elimination of the variables $Y^A$ (see~\cite{Barnich:2006pc,Alkalaev:2009vm} for more details on the ambient space interpretation of the parent system).

\subsection{Global reducibility parameters}

Some important pieces of information on the local gauge theory are captured by the so-called global reducibilty parameters.
Their space is formed by the gauge parameters that give rise to trivial (\textit{i.e.} vanishing) gauge transformations. In the case of reducible gauge theories, this space is to be quotient over the trivial gauge parameters. For a massless spin $s$ field $\phi_{\mu_1\ldots\mu_s}$, the condition implies
$\nabla_{(\mu_1}\lambda_{\mu_2\ldots \mu_s)}=0$ \textit{i.e.} $\lambda_{\mu_1\ldots\mu_{s-1}}$ is a traceless Killing tensor. 
Because the system is irreducible, the space of global reducibilty parameters is in that case given by the space of traceless Killing tensors.
\changed{In the more general setting one can consider higher-order global reducibilty parameters which are parameters of the trivial
gauge transformations for gauge parameters modulo the next order reducibility relations.}

The space of global reducibility parameters is an invariant characteristic of a gauge system, \textit{i.e.} independent of the chosen (equivalent) descriptions. This is known in a rather general context~\cite{Barnich:2001jy} and will be explicitly demonstrated shortly in the case of free systems.
In particular, one can use any equivalent formulation to identify it. For instance, in the case of PM fields it is convenient to use ambient space description based on~\eqref{pmassl} to obtain
\begin{equation}
\label{PMGRP}
\Box \lambda=S \lambda=T \lambda=(U_--1+t)\lambda=\bsd \lambda=(\sd)^t \lambda=0\,.
\end{equation}
The last condition implies that $\lambda$ is polynomial in $X$. The remaining conditions say that
$\lambda$ is described by two row traceless Young tableaux  with rows of length $s-1$ and $s-t$ if one restricts to
the spin-$s$ case by requiring $(N_P-s+t)\lambda=0$.
These properties follow from known facts on representations of the Howe dual pair $\mathfrak{o}(d+2)$ and $sp(4)$.
In terms of intrinsic coordinates on AdS, the global reducibility parameters are determined by the conditions
$(p\cdot \nabla)^t\lambda(x,p)+\ldots=0$ and $\d_p\cdot\,\d_p\lambda=0$, where $\lambda=\lambda_{\mu_1\ldots \mu_{s-t}}p^{\mu_1}\ldots p^{\mu_{s-t}}$. 
A rigorous proof of the isomorphism between these two representations for global reducibilities can be given using the parent formulation.

The global reducibility parameters are nicely described within BRST formalism. Let us for simplicity consider a free system. It can be formulated in the  BRST first quantized terms using the BRST operator $\Omega$ so that the equations of motion and the gauge symmetries read as
\begin{equation}
 \Omega \Phi=0\,, \quad \gh{\Phi}=0\,,\quad \qquad \delta \Phi =\Omega \lambda\,, \quad \gh{\lambda}=-1\,, \qquad \ldots
\end{equation}
If the  BRST operator is proper the space of global reducibility parameters is just the cohomology of $\Omega$ at ghost degree $-1$~\cite{Barnich:2005bn}. In particular, when represented in these terms it is clear that this space is an invariant characterization of the system -- no matter which formulation (among the locally equivalent ones) is being used. In particular, if the space of reducibilities is empty this implies that all the gauge symmetries are St\"uckelberg (algebraic) and the system is equivalent to one without gauge freedom. Analogously, higher global reducibilities are described by $\Omega$-cohomology at ghost degree $-2,-3$ etc. Note that reducibility parameters are directly related to surface charges (see~\cite{Barnich:2005bn} for AdS gauge fields and~\cite{Barnich:2001jy} for the general case) and these notions remain meaningful at nonlinear level as well.

To make contact with the literature, let us mention that in the unfolded formulation of HS fields the module of 1-forms (the so-called gauge module, see \textit{e.g.}~\cite{Vasiliev:2001wa,Bekaert:2005vh}) is known to coincide with the space of global reducibility parameters. It follows from the above BRST cohomology identification that this applies to a general gauge systems.
Indeed, according to~\cite{\BGST,Barnich:2006pc} the module of 1-forms is always given by the cohomology of $\Omega$ at ghost degree $-1$. This extends to higher global reducibilities which are identified with the modules of $p$-forms with $p>1$. 
%Although we now restricted ourselves to free system the notion remains meaningful at the nonlinear level.

\subsection{Reducible multiplet of PM fields}
\label{sec:reducible}

The constrained system~\eqref{pmassl} (or \eqref{pmassl-1}) describes an irreducible field if one further restricts to a particular spin by imposing the condition $(N_P-s)\Phi=0$. We now describe a reducible multiplet of (PM) fields which is directly related to usual depth-$1$ gauge fields.

Consider the following system on the domain $X^2<0$  of $\fR^{d,2}$:
\begin{equation}
\label{reducible}
\begin{gathered}
\bsd \Phi=(U_--2)\Phi=0 \,,\quad (S^2-4T\Box)\Phi=0\,, \quad \Box^{\ell_1}{S}^{\ell_2}{T}^{\ell_3}\Phi=0\,,\\
\Phi\sim \Phi+\sd \lambda
\end{gathered}
\end{equation} 
where $\quad \ell_1+\ell_2+\ell_3=\ell$ and $\lambda$ satisfy the same constraints except that  $U_--2$ is replaced with $U_-$. This system simultaneously describes PM fields of all spins and depths $1,3,\ldots, 2\ell-1$. Again, the truncation to a particular spin $s$ can be achieved by imposing $(N_P-s)\Phi=0$.

To see this let us first consider
the simplest {nontrivial} case of $\ell=2$. The solution of the $\Box^{\ell_1}{S}^{\ell_2}{T}^{\ell_3}\Phi=0$ with 
$\ell_1+\ell_2+\ell_3=2$ can be uniquely written as
\begin{equation}
 \Phi=\Phi_0+\bar T \Phi_1+\bar S\Phi_2 +\bar\Box \Phi_3\,,\qquad \Box\Phi_\alpha=T\Phi_\alpha=S\Phi_\alpha=0\,.
\end{equation} 
Because $\sd,\bsd,U_-$ do not send $\mathfrak{o}(d,2)$-traceless elements to traceful and vice versa the system decomposes into that
for $\Phi_0$ and the rest. The one for $\Phi_0$ is by definition that for usual depth $1$ gauge fields. To see what the remaining fields $\Phi_\alpha$ with $\alpha=1,2,3$ describe let us analyze $\bsd \Phi=0$. This implies
$\Phi_2=-\half\bsd \Phi_3$ and $\Phi_1=\half \bsd\bsd \Phi_3$ so that $\Phi_{1,2}$ are auxiliary fields. To see what does $\Phi_3$ describe observe that its gauge transformation is $\Phi_3\sim \Phi_3+\sd \lambda_3$ and the homogeneity degree $(U_--4)\Phi_3=0$ so by definition $\Phi_3$ is a depth $3$ PM field. Here $\lambda_3$
denotes the respective component of the gauge parameter. 

% A subtle point is that the second constraint in~\eqref{reducible} does not play 
% any role for $\ell=2$. However it starts to do so for higher $\ell$. Without it, 
% already for $\ell=3$ in addition to depth-$5$ PM field the component quadratic 
% in $\bar\Box,\bar S,\bar T$ contains a field of depth $1$. To see how it appears 
% let us decompose quadratic polynomials in $\bar\Box,\bar S,\bar T$ into 
% irreducible components with respect to the algebra $\mathfrak{sp}(2)$ formed by $\sd,U_-,\bsd$. 
% In addition to irreducible 5-dimensional 
% representation with highest weight $4$ (this corresponds to depth $5$ field) 
% there is an $\mathfrak{sp}(2)$-singlet $\bar S^2-4\bar\Box\bar T$. Its coefficient is a 
% depth $1$ gauge field as it follows from analyzing its weight and the gauge 
% transformation.

In the general case it can be shown that the spectrum of PM fields in the
multiplet is in one-to-one correspondence with the spectrum of irreducible $\mathfrak{sp}(2)$ components in
the $\mathfrak{sp}(2)$-module of polynomials in $\bar T,\bar \Box, \bar S$ of order not
higher than $\ell-1$. In other words this is a direct sum of symmetric tensor
powers up to order $\ell-1$ of the standard $3$-dimensional $\mathfrak{sp}(2)$-modules.
To eliminate multiplicities one needs to keep only one irreducible module at each homogeneous component:
the one whose highest weight vector have the form ${\bar\Box}^k\psi_k$ with
$\psi_k$ totally traceless. Multiplicities necessarily contain $\mathfrak{sp}(2)$-invariant element ${\bar S}^2-4\bar \Box \bar T$. Indeed, the irreducible modules of $\mathfrak{sp}(2)%\simeq \mathfrak{o}(3)
$ are realized on $\mathfrak{sp}(2)$-traceless (with
respect to the invariant metric) tensors. Those proportional to ${\bar S}^2-4\bar \Box \bar T$
are hence the multiplicities. But such elements are not present as the condition $S^2-4T\Box$ is contained among
 the constraints~\eqref{reducible}. 
% 
% 
% . Taking this quotient precisely
% eliminate the traces. Finally, taking a quotient with respect to ${{\bar
% \Box}^\ell_1}{\bar S}^{\ell_2}{\bar T}^{\ell_3}$ and ${\bar S}^2-4 \bar \Box
% \bar T$ is equivalent to imposing the dual constraints i.e. the second and the
% third constraint of~\eqref{reducible}.
% 

\section{Boundary values in the (gauge-) covariant approach}\label{Bdyvalues}

The paper started in Section \bref{singletons} {(see also~\cite{Bekaert:2012vt})} with a general discussion of boundary behavior in terms of intrinsic AdS geometry.  Using the simplest example of a massive AdS scalar field, we already introduced the notion of boundary values and discussed various subtleties such as the presence of obstructions to the extension in the bulk.

\subsection{Boundary data for partially massless fields}
\label{sec:PMF-bound}
%%%%%%%%%%%%%%%%%%%%%
According to~\cite{Bekaert:2012vt} the description of the boundary values is 
achieved by considering a constrained system determimining AdS field as defined 
in the vicinity of the hypercone $X^2=0$ (rather than on the hyperboloid 
$X^2=-1$) and interpreting it in terms of homogeneous fields defined on the 
hypercone. {Here we explicitly concentrate on boundary values with 
shadow-type asymptotic behaviour and hence employ the equation~\eqref{pmassl-1} or 
\eqref{pmassl} where the homogeneity degree is fixed by $\Delta_-=t+1-s$. For 
definiteness we use~\eqref{pmassl-1}. The current-type boundary values are 
obtained in exactly the same way using a different constrained system involving 
$\Delta_+$ in the homogeneity constraint (see Section~\bref{sec:current}).}

This prescription can be made fully gauge-covariant by using the BRST 
operator~\eqref{PM-BRST}. In this form the  prescription of boundary values is 
understood as a map that sends a gauge theory (with unconstrained gauge 
parameters) on AdS space  to another gauge theory on the conformal boundary 
whose gauge parameters are also unconstrained. However, even this form of the 
map is somewhat implicit because the same $\Psi$ encodes fields defined on the 
hyperboloid and their boundary values defined on the projective hypercone. In 
order to make it explicit, one passes to the parent description.

In parent terms, the map that sends a bulk gauge theory to the boundary one 
simply amounts to considering the {generating function} $\Psi$  to be defined on the boundary and 
replacing the AdS connection $\omega$ and compensator $V$ with their conformal 
counterparts in~\eqref{constr-inter} so that the system is determined by the 
BRST operator of the same structure \begin{equation} \label{conf-parent} 
\Omega_{conf}=\nabla+Q\,, \qquad \Psi=\Psi(x^a,\theta^a|Y,P,b)\,, \end{equation} 
where $\Psi$ is subject to the same target space constraints, the variables 
$x^a,\theta^a$ denote the coordinates on the conformal boundary and their De 
Rham differentials, and $\nabla$ and $V$ are the conformal connection and the 
compensator. In this way, the map indeed sends a gauge field theory defined on 
the hyperboloid to a gauge field theory defined on its conformal boundary. 

Let us spell out explicitly the structure of the conformal compensator and covariant derivative.
In a suitable local frame $e_\pm,e_a$ one can assume $V^+=1,V^-=V^a=0$ and (see~\cite{Bekaert:2009fg,Bekaert:2012vt} for more details)
\begin{equation}
  \nabla=\derham-\omega^a_{b} (y^b\dl{y^a}+p^b\dl{p_a})-e^a[(Y^++1)\dl{y^a}- y_a \dl{Y^-}+P^+\dl{p^a}-p_a\dl{P^-}]\,.
\end{equation} 
Adjusting the frame and local coordinates one can in addition put $\omega=0$ and $e^a=\theta^a$.

To make contact with the considerations of~\cite{Bekaert:2012vt}, we observe that the gauge $A=0$ admissible
for the bulk system~\eqref{constr-inter} is {also} admissible for
the boundary system~\eqref{conf-parent} by the same reasoning. In this gauge \eqref{constr-inter} reduces to the conformal version of
\eqref{parent-gf}-\eqref{parent-gf-gs}, \textit{i.e.} the same system but with $F$ depending on $x^a$ and AdS $\omega,V$ replaced with their conformal counterparts. For $t=1$ this is precisely the equations of motion and gauge tranformations for boundary values obtained in~\cite{Bekaert:2012vt}. We have thus confirmed that going to the boundary commutes with the partial gauge condition employed in~\cite{Bekaert:2012vt}.

The following important remark is in order. Although we extensively use BFV-BRST 
formalism to describe gauge theories the involved BRST differentials are not 
necessarily proper in the sense that they do not always take into account all 
the gauge symmetries of the equations. For instance, the above parent BRST 
operator for PM fields is proper while its conformal counterpart describing 
boundary values is in general not. Indeed, taking the scalar field $s=0$ as an 
example and assuming $d\neq 2\ell$ the BRST operator for boundary values describes off-shell system 
with trivial equations and no gauge symmetries. In the standard considerations such BRST operator is regarded as non proper as it does not take into account the gauge symmetries (e.g. shifts by an arbitrary parameter)
present in the system. This supports the point of view that not only proper BRST differentials are of 
primary importance.

To complete the discussion of the general aspects of boundary values let us note 
that the way we describe boundary values is directly analogous to obtaining the 
Hamiltonian formulation which, from this perspective, can be seen as a theory of 
boundary values of fields with the boundary being the surface of initial data. 
To illustrate this point, in Appendix~\bref{sec:Minkowski} we consider a toy model 
of the Klein-Gordon field on flat spacetime and show that the parent approach to boundary values
indeed reproduces the Hamiltonian phase space for the Klein-Gordon field.

\subsection{Boundary values in terms of components}
\label{sec:bound-val-exp}

We now explicitly analyze the partially gauge fixed system describing the boundary values. For this we find it more convenient to work with the formulation based on~\eqref{pmassl}. The  respective counterpart of~\eqref{parent-gf} reads explicitly as
\begin{equation}
 \nabla F=0\,,
 \quad \Box F=S F=T F=\bsd F=\big((Y+V)\cdot \d_Y+\Delta\big)F=0\,,
\end{equation}
where $F=F(x|Y,P)$ and $\Delta=t+1-s$. The gauge parameters $\lambda$ satisfy the same equations but with
$\Delta=1-s$ and the gauge law is $\delta F=(\sd)^t \lambda$.

Constraints $\nabla,(Y+V)\cdot \d_Y,\bsd$ uniquely fix the dependence of $F$ on $y^a,Y^+,P^+$ in terms of
$\phi(x|u,w,p)=F|_{y^a=Y^+=P^+=0}$ where $u:=Y^-$ and $w:=P^-$. 
If $\phi$ describes a spin-$s$ PM field, then $(p^a\dl{p^a}+w\dl{w}-s)\phi=0$ and the remaining equations take the form (these are direct generalizations of Eqs. (5.12)-(5.14) from~\cite{Bekaert:2012vt})
\begin{align}
\tilde\Box \phi+\dl{u}\left(d-2\Delta-2u\dl{u}\right)\phi=&0\,,\label{eq1}\\
\left(\dl{p}\cdot\dl{x}\right)\phi +\dl{w}
\left(d-\Delta-1+s-2u\dl{u}-w\dl{w}\right)
\phi=&0\,,\label{eq2}\\
\left(\dl{p}\cdot\dl{p}\right)\phi-2u\left(\dl{w}\right)^2\phi=&0 \label{eq3}\,,
\end{align}
where $\tilde\Box=\Box_0+2(p\cdot\dl{x})\dl{w}+p^2(\dl{w})^2$. The equations corresponding to the gauge parameter 
are obtained for $\Delta=1-s$ and $s$ should be replaced by $s-t$.

Let us analyze what these equations encode for the PM field itself, \textit{i.e.} for $\Delta=1+t-s$.
At $u=0$ the coefficient in~\eqref{eq2} never vanishes
because $t\leq s$. Just like in the massless case for $d$ even the general solution is parametrized by two functions of $x$ and $p$:
$\phi_0^0:=\phi|_{u=w=0}$ and $\psi_\ell$, where the latter enters $\phi$ as $u^\ell\psi_\ell$ {with}
$\ell=\half(d-2\Delta)=\frac{d-2}{2}+s-t$. Both are subject to some conformal equations, as explained below.

Repeating the analysis of \cite{Bekaert:2012vt} in the present case one finds
equations for $\phi_0:=\phi|_{u=0}$ encoded in those for $\phi(x,p,w)$
\begin{equation}
\begin{gathered}
\label{FT+pgc}
 \tilde\Box^\ell
 %{\frac{d-2}{2}+s-t}
 \phi_0=0\,,
 \\
 \left(\dl{p}\cdot\dl{x}\right) \phi_0+\dl{w}
 %(d-2+2s-t-w\dl{w})
 (2\ell+t-\dl{w})
 \phi_0=0\,, \quad
 \left(\dl{p}\cdot\dl{p}\right)\phi_0=0\,.
\end{gathered}
\end{equation}
Just like in the massless case considered in~\cite{Bekaert:2012vt} the second 
equation determines $\phi_0$ in terms of $\phi_0^0=\phi_0|_{p_-=0}$ while the 
third equations says that $\phi_0^0$ is traceless. At the same time the first 
equation imposes differential equations on $\phi_0^0$. More precisely, 
$(\tilde\Box^{\frac{d-2}{2}+s-t}\phi_0)|_{w=0}=0$ is in fact invariant under 
gauge transformations with unconstrained parameter while the remaining equations 
encoded in $\tilde\Box^{\frac{d-2}{2}+s-t}\phi_0=0$ are partial gauge 
conditions. Note that the conformal equations encoded in~\eqref{FT+pgc} fit into 
the classification~\cite{Shaynkman:2004vu} and also belong to the class of 
conformal gauge fields described in~\cite{Vasiliev:2009ck}.

As for equations for $\psi_\ell(x,p)$ one finds the partial conservation condition
\begin{equation}
 \left(\dl{p}\cdot\dl{x}\right)^t\psi_\ell=0\,,
\end{equation}
in agreement with~\cite{Dolan:2001ih}.

As an example let us consider the case $d=4$ and $t=s$ (\textit{i.e.} maximal depth) where the equations are of second order.
The first equation at $w=0$ gives
\begin{equation}
\label{d4-s2}
 \Box_0\phi_0^0-\frac{2}{3} (p\cdot \dl{x})(\dl{p}\cdot \dl{x})\phi_0^0+\frac{1}{6}p^2(\dl{p}\cdot \dl{x})(\dl{p}\cdot \dl{x})\phi_0^0=0
\end{equation}
To obtain the explicit form of gauge transformations one repeats the analysis for the gauge parameter. In terms of $\lambda^0_0(x)$ (in the present case $s=t$,
$\lambda$ is independent of $p$) the gauge transformation for $\phi_0$ read as
\begin{equation}
 \delta \phi_0^0=\big((p^a\dl{x^a})^2-\frac{1}{4}p^ap_a \Box_0\big)\lambda^0_0\,.
\end{equation}
The gauge invariance of~\eqref{d4-s2} can be easily checked directly. Note that these are precisely the conformal spin $2$ equations from~\cite{Deser:1983mm} {(see also~~\cite{Deser:2004ji})} which, thanks to conformal invariance, can also be seen as equations on $AdS_4$. More precisely, equations describing $s=t=2$ PM fields. Analogous remark applies to $s=t=1$ case.

{
More generally, the higher-depth FT equations are of order $d-2+2(s-t)$ in the derivatives. One can notice that in $d=4$ and for maximal depth $t=s$ they are of second order. Again, because of conformal invariance they can be seen as second order equations for gauge fields on $AdS_4$ with a gauge law of order
$t$ in derivatives. However, these conformal fields \changed{do} not seem to coincide  with PM fields of maximal depth $t=s$ in $AdS_4$ for $s>2$.}~\footnote{We thank E. Joung and K.~Mkrtchyan for a useful discussion of related issues.} We do not discuss (higher-depth) FT equations here in more details because later in Section~\bref{sec:manifest} we present the gauge-invariant and manifestly conformal form of them.

\subsection{Relation to the unfolded formulation}

If one does not want to use partial gauge conditions a possible way to study the gauge invariant system~\eqref{conf-parent} is to reduce 
it further to the unfolded formulation and compare with the known unfolded forms of conformal gauge fields~\cite{Vasiliev:2009ck}. 
According to the general prescription of~\cite{\BGST,\BGadS,\BGnlp} the unfolded formulation can be arrived at 
by eliminating a maximal amount of purely target space contractible pairs. 
In the case at hand this implies reduction to the cohomology of $Q=\sd\dl{b}$ \textit{i.e.} the target space part of the total BRST operator.

Because there is only one target space ghost variable left (ghost momenta $b$) the $Q$ cohomology can be found at target space ghost degree $-1$ and $0$. For degree $-1$ the coboundary condition is missing and the cohomology can be identified with the subspace of elements of the form $b\,\phi(Y,P)$ with $\phi$ satisfying
\begin{equation}
 \Box\phi=S\phi=T\phi=(\bsd)^t\phi=(U_--t+1)\phi=\sd\phi=0\,,
\end{equation}
so that $\phi$ can be represented as polynomial $\phi(Y+V,P)$ with coefficients associated to totally traceless $2$-row Young tableaux 
with 1st row of length $s-1$ and 2nd of length $s-t$.  So that at degree $-1$ one has exactly the same space as for AdS PM fields. This is a general feature: the unfolded formulation of boundary values with $\Delta_-$
asymptotic has the same 1-form sector as its AdS counterpart. Similar observation is known in the unfolded
approach~\cite{Vasiliev:2012vf} to boundary dynamics. {More generally, the sector of $1$-forms coincides with the 
space of inequivalent reducibilities as is clear from the cohomological characterization of both spaces.}

Let us now turn to the degree 0 sector. The cohomology can be identified with the following quotient space:
\begin{equation}
\label{WM}
 \phi \simeq \phi+\sd \chi\,,\quad  \Box\phi=S\phi=T\phi=\bsd\phi=(U_--2)\phi=0\,,
\end{equation}
where $\phi=\phi(Y,P)$ and for simplicity we have restricted to $t=1$ case. In
fact, we have implicitly described this space in~\cite{Bekaert:2012vt}. Although it is
difficult to explicity solve the constraints in terms of $Y$-variables,
the solution can be explicitly characterized in terms of the auxiliary
$d$-dimensional space by allowing $\phi$ to depend on extra coordinates
$x^a$ and uniquely fixing this dependence through $\nabla \phi=0$. For this to
be always well-defined we take formal power series in $x^a$ in contrast to the
considerations of Section~\bref{sec:bound-val-exp} and~\cite{Bekaert:2012vt}. In spite of this difference all the
computation remain valid and we conclude that this space is nothing but the
space of solutions to (higher-depth) gauge fixed FT equations in the
space of formal series in $x^a$. In these terms, the equivalence relation in~\eqref{WM} takes a quotient with respect to the residual gauge symmetry. One concludes that the $0$-form sector is indeed that of FT
fields.

\subsection{Current-type asymptotics}
\label{sec:current}
In the partially massless case the dual choice of the asymptotic behaviour is to take $\Delta=\Delta_+=d-\Delta_-=d+s-1-t$.  Just like in the massless case the implementation of this asymptotic behavior in a gauge invariant and manifestly $\mathfrak{o}(d,2)$-invariant way requires modification of the constraints by the automorphism~$P\to \dl{P}, \quad \dl{P}\to -P$ \cite{Bekaert:2012vt}. 
In addition, one employs a different subalgebra of constraints
to generate gauge transformations (in terms of the BRST description this amounts to shifting the ghost degree). 
The simple rule is that one changes the representation for all the ghosts except those associated to the constraints of vanishing homogeneity in $P$ (\textit{i.e.} $U_-,\Box$ in our case).

Starting with~\eqref{pmassl} one ends up with
\begin{equation}
\label{pmassl-curr}
\begin{gathered}
\Box\Phi=(N_X+d+s-t-1)\Phi=(S)^t \Phi=0\,,\\
\Phi \sim \Phi+\sd \chi_1+\bar S \chi_2+\bar T\chi_3\,,
\end{gathered}
\end{equation}
where the parameters $\chi_\alpha$ are subject to the gauge parameter version of the same constraints 
{(see~\cite{Bekaert:2012vt} {for general discussion of constraints on gauge parameters)}}. When considered on the domain $X^2<0$ of $\amb$ this system of constraints is equivalent to~\eqref{pmassl}. The same applies to the parent formulations based on~\eqref{pmassl-curr}.

According to the general prescription, the boundary values associated with current-type boundary behavior
are described by the same constrained system but defined in the vicinity of the hypercone rather than
the hyperboloid. Expanding the constraints around the point $X^+=1,X^-=x^a=0$ (a rigorous argument can be done using the parent formulation where compensator $V^A$ plays a role of a point around which the expansion is done) one finds
that the first two constraints eliminate the dependence on $X^+,X^-$. The 4th and 5th allow to gauge away
$P^+$ and $P^-$ while the last one allows to assume the field traceless. 
Finally the remaining 3rd constraint imposes the partial conservation condition 
$(\dl{x^a}\dl{p_a})^t\phi^0_0(x,p)=0$.

Alternatively one can start with the constrained system~\eqref{pmassl-1}. In 
this case, the constraints $(S)^t$ and $\bar S$ in~\eqref{pmassl-curr} are 
respectively replaced by $S$ and $(\bar S)^t$. To see that the 
{resulting} system describes partially conserved currents one modifies the above 
argument as follows: gauge transformation generated by $(\bar S)^t$ allows to 
get rid of the dependence on $(P^-)^l$ with $l\geq t$ only. In such a subspace 
the remaining constraint {$S\Phi=0$
%$S\Phi(xp)=(\dl{P^-}(\cdot)+\dl{x^a}\dl{p_a})\Phi(P^-,x,p)=0$ 
implies 
$(\dl{x}\cdot\dl{p})^t\phi^0_0=0$% where $\Phi_0=\Phi|_{P^-=0}$. 
%Here, $(\cdot)$  denotes some nonvanishing coefficients originating from $\dl{X^+}$
% and the constraint fixing the homogeneity degree.
}

{Let us stress that boundary values with $\Delta_+$ (\textit{i.e.} current-type) 
asymptotics are not gauge fields. This can be traced to the fact that the bulk 
gauge transformations become St\"uckelberg near the boundary and hence do not 
determine genuine (not equivalent to algebraic) gauge transformations of the 
boundary values.}

To complete the discussion of boundary values in the gauge covariant approach, in 
Appendix~\bref{sec:GI} we show how the formulation of boundary values can be 
obtained in terms of curvatures using spin-$1$ field as a simple example.

\subsection{Massive fields}
% The ambient analysis of boundary values of a (massive) scalar and Fronsdal fields on AdS has been performed
% in~\cite{Bekaert:2012vt}. Here we address slightly more involved partially massless and massive fields (perhaps mixed symmetry fields?)

For a masive spin $s$ field an ambient formulation is based on essentially the same constrained system
(see~\cite{Grigoriev:2011gp,Alkalaev:2011zv} for more details)
\begin{equation}
 T\Phi=S\Phi=\Box\Phi=\left(X\cdot\dl{X}+\Delta\right)\Phi=0\,,\quad \qquad \Phi\simeq \Phi+\sd  \chi
\end{equation}
where $\Delta$ is generic (in particular, $\Delta\neq d/2-\ell$ with positive integer $\ell$). 
In this case the gauge symmetry can be fixed by imposing $\bsd  \Phi=0$ as a gauge condition. 
Equations are again the same except for the coefficients involving $\Delta$. 
Because $\Delta$ is generic none of them vanishes and the boundary value is an off-shell traceless field of weight $\Delta$.

\section{Manifestly conformal formulations}
% of (generalized) FT equations
% and partially conserved currents}
\label{sec:manifest}

\subsection{Higher-order singletons}
\label{higher-ordersinglet}

In terms of the ambient space the higher-order singleton can be described as follows.
Consider the following equations for an ambient space scalar
\begin{equation}
\label{eq-l-scalar}
 \Box^\ell \Phi=0\,, \qquad (H+\ell+1)\Phi=0\,,
\end{equation}
where $H=-X\cdot \d_{X}-\frac{d+2}{2}$. Note that $E=\Box,F=-\bar\Box$ and $H$ satisfy the standard $\mathfrak{sp}(2)$-relations.  In particular, 
$H=\commut{E}{F}$.

% Here and below we use the following notations
% \begin{equation}
%  \Box=\half \dl{X} \cdot \dl{X}\,, \qquad \bar\Box=\half X\cdot X\,, \qquad h=\commut{\Box}{\bar\Box}=X\cdot \dl{X}+\frac{d+2}{2}
% \end{equation}
% Note also the remaining $\mathfrak{sp}(2)$ relations $\commut{h}{\bar\Box}=2\bar\Box$ and $\commut{h}{\Box}=-2\Box$

For $\ell$ a positive integer, one finds the following gauge symmetry
\begin{equation}
 \delta_\lambda \Phi=\bar\Box \lambda
\end{equation}
where $\lambda$ is subject to
\begin{equation}
 \Box^\ell \lambda=0\,, \qquad (H+\ell-1)\lambda=0\,.
\end{equation}
The consistency can be easily checked using
  \begin{equation}
  \commut{\Box^\ell}{\bar\Box}=-\Box^{\ell-1}\ell(H+\ell-1)
 \end{equation}
which is~\eqref{sp2} with $t=\ell$.
The gauge symmetry is right  enough to assume the field to be defined on the hypercone.
Using the homogeneity condition allows one to restrict the ambient field $\Phi$ to the conformal space $\manX_d$ (seen as a section of the hypercone)
and obtain a conformal scalar field $\phi_0$ of weight $\Delta=\frac{d}{2}-\ell$.
Finally, using suitable coordinates on the conformal space the first equation in~\eqref{eq-l-scalar}
implies the polywave equation
\eqref{polywave}.
%$(\dl{x^a}\dl{x_a})^\ell\phi=0$ 
%\textit{i.e.} the conformal equation for a 

Although the complete set of constraints does not form a Lie algebra, it is not difficult to find a closed expression for the BRST
operator~\footnote{In fact, this set of constraints is of the type shown in~\cite{Buchbinder:2007au} to admit a closed expression for the BRST charge.
Furthermore, the constraints determining PM fields and higher depth FT fields also belong to this class.}. Introducing the Grassmann odd ghost variables $c_0,c_1,c_{-1}$ and their conjugate ghost momenta $b_0,b_1,b_{-1}$ the Weyl symbol of the BRST operator reads as
\begin{multline}
\label{HOS-BRST-Weyl}
 \Omega=c_1 \left(\frac{P^2}{2}\right)^\ell+c_0 (P\cdot X)+ c_{-1}\frac{X^2}{2}
 +\\
 +c_0(2c_{-1}b_{-1}-2\ell c_1b_1)
 +\ell c_1 c_{-1}\left(\frac{P^2}{2}\right)^{\ell-1}b_0\,.
\end{multline}
It satisfies the quantum nilpotency  $\Omega*\Omega=\half\qcommut{\Omega}{\Omega}=0$, where $*$ denotes the Weyl star product determined by 
the graded commutators $\qcommut{X^A}{P_B}=\delta^A_B$ and $\qcommut{c_{\alpha}}{b_{\beta}}=\delta_{\alpha\beta}$.

The explicit expression for the operator is obtained by representing the canonical pairs $(X,P)$, $(c_1,b_1)$, and
$(c_0,b_0)$ in the coordinate representation while $(c_{-1},b_{-1})$ in momentum representation and using the Weyl (symmmetric)
ordering. In particular the term proportional to $c_0$ is
\begin{equation}
-N_X-\frac{d+2}{2}+\ell+1-2\ell c_1\dl{c_1}-2b_{-1}\dl{b_{-1}}
\end{equation}
so that the constraint $H+\ell+1$ is reproduced for ghost-independent states. 
For $\ell=1$, the expression \eqref{HOS-BRST-Weyl} reduces to the standard BRST operator for the algebra $\mathfrak{sp}(2)$. 
Note that $\Omega$ does not have quantum corrections: it satisfies both the quantum $\qcommut{\Omega}{\Omega}=0$ and the classical $\pb{\Omega}{\Omega}=0$ master equations, where $\pb{}{}$ denotes the Poisson bracket associated to the $*$-product. Indeed, the quantum corrections to the Poisson bracket contain $3,5,7\ldots$ derivatives {of each argument} and the explicit check shows that these do not contribute to $\qcommut{\Omega}{\Omega}$.

\bigskip

%\subsubsection{Second-order version}
An alternative description of the higher-order singleton can be constructed using the following ambient system (c.f. Subsection \bref{subsambound}):
\begin{equation}
\label{eq-2-scalar}
 \Box \Phi=0\,, \qquad (H+\ell+1)\Phi=0\,.
\end{equation}
For $\ell$ positive integer one finds the following gauge symmetry
\begin{equation}
 \delta_\lambda \Phi=(\bar\Box)^\ell \lambda
\end{equation}
where $\lambda$ is subject to
\begin{equation}
 \Box \lambda=0\,, \qquad (H-\ell+1)\lambda=0\,.
\end{equation}
To check the consistency the following relation is useful
 \begin{equation}
  \commut{\Box}{\bar\Box^\ell}=-\bar\Box^{\ell-1}l(H-\ell+1)\,,
  \label{bbbar}
 \end{equation}
which is~\eqref{sp2} with $t=\ell$.

For the scalar singleton ($\ell=1$) the description based on \eqref{eq-2-scalar} and the one based on \eqref{eq-l-scalar} explicitly coincide. In this case an ambient description of the singleton has been originally proposed in~\cite{Marnelius:1978fs} (see \textit{e.g.}~\cite{Bekaert:2009fg} for more details). Note also that the constrained system with constraints $X^2,X\cdot P,P^2$
underlies the two-time physics approach~\cite{Bars:2000qm}.

It is less trivial to see that the two constructions are equivalent for all 
integers $\ell>0$. As a possible argument, one can repeat the analysis 
of~\cite{Bekaert:2012vt} (alternatively, Section~\bref{sec:bound-val-exp} for 
$s=0$ and $\Delta=\frac{d}2-\ell$). One finds that the space of solutions 
to~\eqref{eq-2-scalar} has a submodule of subleading solutions. These are 
elements that are proportional to $(X^2)^\ell$ (see Section~\bref{subsambound}).

An heuristic explanation of the close relation between the two constructions is the Weyl-algebra automorphism $X\to -\dl{X}$, $\dl{X}\to X$ which relates the $\mathfrak{sp}(2)$ generators via the Chevalley automorphism: $H\to -H$ 
and it interchanges the remaining two $E \leftrightarrow F$. For the latter two, one also exchanges the constraint which is imposed and the one {that generates gauge transformations}, which is the reason why the homogeneity degree of the scalar fields is the same though the generator $H$ changed sign.

\subsection{Higher-depth Fradkin--Tseytlin fields}
%\subsubsection{second-order version}

The identification of (higher-depth) {FT}
gauge fields as boundary values of (partially) massless fields in the bulk gives a manifestly
conformal description of them. However, the ambient space system for boundary
values simultaneously describes both the higher-depth FT fields and the associated partially
conserved currents. Although it is easy to describe the former fields in terms of
components as we did in Section~\bref{sec:PMF-bound} (depth-$1$ case was
considered already in~\cite{Bekaert:2012vt}) the manifestly $\mathfrak{o}(d,2)$-covariant description
is not so straightforward.

We begin with the purely ambient picture. 
According to the general prescription, the boundary data for PM fields is described by the same ambient constraints but considered in the vicinity of the hypercone.
In terms of the generating function $\Phi(X,P)$, the constraints~\eqref{pmassl-1} take the form
\begin{equation}
\label{nnf}
 \Box\Phi=S\Phi=T\Phi=\bsd\Phi=(N_X +t+1-s)\Phi=(N_P-s)\Phi=0
\end{equation}
where we have explicitly restricted to the spin-$s$ field and we are assuming $d$ to be even.
At the moment we do not take into account the gauge invariance.

Let us quotient the space of field configurations over the subspace of configurations of the form  $\bar\Box^{\ell}\lambda$ where $\ell={\frac{d-2}{2}+s-t}$ and $\lambda$ satisfies
\begin{equation}
\label{lambda-const}
  \Box\lambda=S\lambda=T\lambda=\bsd\lambda=
 (N_X+d+s-t-1) \lambda=(N_P-s)\lambda=0\,.
\end{equation}
Note that the homogeneity degree in $X$ is such that \eqref{lambda-const} is compatible
with the constraint $\Box\Phi=0$, as can be seen from the commutation
relation \eqref{bbbar}. It is easy to see that $\bsd \bar\Box^\ell\lambda=0$
while the constraints $S\bar\Box^\ell\lambda=T\bar\Box^\ell\lambda=0$ are fulfilled thanks to the algebra.
 The remaining constraint in~\eqref{nnf} is also fulfilled thanks to
\begin{equation}
 N_X (\bar\Box)^{\ell} \lambda=(\bar\Box)^{\ell}(N_X+2\ell)\lambda=(s-1-t)(\bar\Box)^{\ell}\lambda\,.
 \end{equation}
This factorization can be seen as an extra gauge transformation $\delta \Phi=
\bar\Box^{\ell}\lambda$.

The usual gauge transformation is $\delta_\alpha \Phi=(\sd)^t \alpha$, where $\alpha$ takes values in the space
of gauge parameters:
\begin{equation}
\label{Fgtransf}
 \Box\alpha=S\alpha=T\alpha=\bsd\alpha=(N_X-s+1)\alpha=(N_P-s+t)\alpha=0\,,
\end{equation}
We now quotient this space over the subspace of elements of the form $\bar\Box^{\ell+t}\beta$ with $\beta$
satisfying
\begin{equation}
  \Box\beta=S\beta=T\beta=\bsd\beta=
 (N_X+d+s-1) \beta=(N_P-s+t)\beta=0\,.
\end{equation}
The consistency check is a direct analog of that for the space of field configurations.

It turns out that $(\sd)^t$ determines a well-defined map from equivalence classes of gauge parameters $\alpha$
to equivalence classes of field configurations $\Phi$ and hence the respective gauge theory is well-defined.
Indeed, gauge parameter $\alpha=\bar\Box^{\ell +t}\beta$ gives rise to the field configuration of the form $\bar\Box^\ell \lambda$ as
\begin{equation}
%$
 ({\sd})^t\bar\Box^{\ell +t}\beta=\bar\Box^\ell \lambda(\beta)\,,
% $
\end{equation}
for some $\lambda(\beta)$. Moreover,  $\lambda(\beta)$ satisfies~\eqref{lambda-const}. To see this one acts by $\Box$
on both sides of the above relation and gets zero in the left hand side and $\bar\Box^\ell \Box\lambda(\beta)$ in the right hand one. Because $\bar\Box^\ell$ has trivial kernel $\Box \lambda(\beta)=0$. Similarly, acting with $\bsd$ gives $\bsd \lambda(\beta)=0$ so that $T\lambda(\beta)=S\lambda(\beta)=0$ thanks to the algebra. Consider as an example
the case $t=1$:
\begin{equation}
 \sd (\bar\Box^{\ell+1}\beta)=\bar\Box^\ell \lambda(\beta)\,, \qquad
 \lambda(\beta)=(\ell+1)\bar S \beta+\bar\Box \sd \beta\,.
\end{equation}
That $\lambda(\beta)$ is indeed annihilated by $S,\bsd, T,\Box$ as well as $N_X+d+s-2$
can be checked directly.

The consistent factorization described above eliminates the partially-conserved current
configurations from the boundary data encoded by constraints~\eqref{nnf}. To
see this let us turn to the parent formulation where $X^A$ is replaced with $Y^A+V^A$.
In the frame where $V^+=1, V^-=V^a=0$ the constraint $(\bar\Box)^\ell=(Y^-)^\ell+\ldots$
and taking the quotient with respect to this constraint eliminates the conserved
current configurations as according to the analysis of Section~\bref{sec:PMF-bound} (see also~\cite{Bekaert:2012vt})
such configurations enter the formulation as $(Y^-)^\ell$. 
%In the case of scalar this is almost obvious.
\vspace{2mm}

Let us also present the corresponding description in the manifestly local and fully gauge covariant formulation 
using the parent BRST approach. 
To start, consider the space of polynomials in the variables $P$ and in the ghost $b, \,\,\gh{b}=-1$ with coefficients that are formal power series in $Y$. 
{The target space BRST operator is $Q=(\sd)^t \dl{b}$.}
Let $\cH$ denote the {subspace} singled out by
%\begin{equation}
  \begin{gather}
  \label{b-const}
\Box\Psi=S\Psi=T\Psi=\bsd\Psi=(N_P-s+t\, b\d_b)\Psi=0\,,\\
\label{2b-const}
\big((Y+V)\cdot \d_Y-s +1+t-t\,b\d_{b}\big)\Psi=0\,.
\end{gather}
%\end{equation}
The conditions~\eqref{nnf} and \eqref{Fgtransf} for respectively $\Phi$ and $\alpha$ are reproduced by taking $\Psi=\Phi+b\alpha$.

Let $\cH_0$ be a quotient of $\cH$ modulo the subspace of elements of the form $\bar\Box^{\ell+t\,b\d_b} \Lambda$
where $\Lambda$ satisfies~\eqref{b-const} and the following modified version of~\eqref{2b-const}:
$\big((Y+V)\cdot \d_Y+d+s-1-t+t\,b\d_{b}\big)\Lambda=0$ so that for $\Lambda=\lambda+b\beta$ the respective conditions on $\lambda,\beta$ are reproduced.

It is straightforward to check that $Q$ determines  a well-defined operator on the quotient space $\cH_0$, which we keep denoting by $Q$. 
Finally, the manifestly local and conformal formulation of the higher-depth FT fields is given by the following parent BRST operator
%\begin{equation}
 $\Omega=\nabla+Q\,$.
%\end{equation}
The string field $\Psi(x,\theta)$ takes values in the space $\cH_0$. 
The $\mathfrak{o}(d,2)$-invariance is built in the construction 
because $\cH_0$ is defined in terms of the $\mathfrak{o}(d,2)$-invariant generators of $\mathfrak{sp}(4)$ while $\nabla$ is a flat
$\mathfrak{o}(d,2)$-connection. In particular the $\mathfrak{o}(d,2)$ transformation with parameter $f^A_B$ acts on the fields according to $F^A_B(x)J_A^B $, where $J^A_B$ denotes a realization of $\mathfrak{o}(d,2)$ on $\cH$ and $F^A_B(x)$ is a covariantly constant extension of $f^A_B$.

{It would be interesting to investigate the relation of these generalized FT equations for higher-depth with the higher-derivative Fronsdal-like equations and actions of \cite{Joung:2012qy}.}

\subsection{BRST operator for higher-depth Fradkin--Tseytlin fields}

Let us finally present even more transparent form for the system of the previous section, where the factorization is implemented through the BRST operator as an extra gauge invariance. To this end let us extend $\cH$ by extra ghosts $\pi$ and modify~\eqref{2b-const} as follows
\begin{equation}
\label{Ybp}
 \left((Y+V)\cdot \d_Y-s +1+t-tb\d_{b}+2(\ell+tb\dl{b})\pi\dl{\pi}\right)\Psi=0\,,
\end{equation}
This constraint determines the homogeneity in $Y+V$ of elements with definite ghost dependence. In particular,
for the homogeneity of $\psi_0,b \psi_b,\pi\psi_\pi,b\pi \psi_{b\pi}$ one respectively gets
\begin{equation}
\begin{gathered}
 %(Y+V)\cdot \d_Y
 -s+1+t\,,
 %)\phi_0=0\,,
 \qquad
 %(Y+V)\cdot \d_Y
 -s+1\,,
%\quad
 %)\phi_b \,,
 %\\
 \qquad
 %(Y+V)\cdot \d_Y+
 d+s-1-t\,,
 %)\phi_\pi \,,
 \qquad
 %(Y+V)\cdot \d_Y+
 d+s-1\,.
 %)\phi_{b\pi}
\end{gathered}
\end{equation}

Because the constraints are nonlinear the respective BRST operator should involve (higher) structure operators and could be of higher rank.  However as we only have two ghosts and both are fermionic it can only be cubic in ghosts.
Moreover, it can even be found explicitly:
\begin{equation}
 Q^\prime=(\sd)^t(1-\pi\dl{\pi})\dl{b}+(\bar\Box)^{\ell+t\,{b\dl{b}}}\dl{\pi}+F_t\pi\dl{\pi}\dl{b}\,,
\end{equation}
where the operator $F_t$ is defined as follows $(\sd)^t\bar\Box^{\ell+t}=\bar\Box^\ell F_t$. Its explicit form can be found recursively. In particular for $t=1$ one finds
\begin{equation}
 F_1=(\ell+1)\bar S+\bar\Box \sd\,.
\end{equation}
The respective manifestly conformal parent formulation is also immediately obtained by taking
$\Omega^\prime=\nabla+Q^\prime$ and $\Psi=\Psi(x,\theta|Y,P,b,\pi)$ where $\Psi$ is subject to
\eqref{b-const} and \eqref{Ybp}. 

%In this case the constraints
% {for $\Psi=\Psi(x,\theta|Y,P,b,\pi)$} take the form
% \begin{gather}
% \Box\Psi=S\Psi=T\Psi=(N_P-s+t b\d_b)\Psi=0\,,\qquad (\bsd)^{t-(t-1)\pi\d_\pi}\Psi=0\,,\\
% ((Y+V)\cdot \d_Y-s +1+t-b\dl{b}+2(\ell+b\dl{b})\pi\dl{\pi})\Psi=0\,.                                                                                           \end{gather}
% The BRST operator reads as {$\Omega^{\prime\prime}=\nabla+Q^{\prime\prime}$ with}
% \begin{equation}
%  Q^{\prime\prime}= \sd(1-\pi\dl{\pi})\dl{b}+(\bar\Box)^{\ell+{b\d_{b}}}\dl{\pi}+((\ell+1)\bar S+\bar\Box \sd)\pi\dl{\pi}\dl{b}\,.
% \end{equation}
% Note that in the formulation based on~\eqref{pmassl-1} the BRST operator is explicit. This is in agreements with~\cite{Alkalaev:2009vm,Alkalaev:2011zv} where the formulation of the same type has been used to describe mixed-symmetry PM fields in a concise way.

To complete the discussion of $\mathfrak{o}(d,2)$-covariant formulations of higher-depth FT fields let us mention that, just like
the order-$\ell$ singleton can be described  in two equivalent ways~\eqref{eq-l-scalar} and \eqref{eq-2-scalar},
a spin-$s$ depth-$t$ FT field also admits a second description based on $\Box^\ell\Phi=0$ with
$\ell={\frac{d-2}{2}+s-t}$ and the gauge symmetry generated by $\bar\Box$. 
However, compatibility with the gauge invariance requires to adjust the remaining constraints.  
The resulting set of constraints is related to the one considered above by an automorphism of the algebra $\mathfrak{sp}(4)$.
Note also that an alternative manifestly conformal description can be constructed using constraints~\eqref{pmassl-1} in place of~\eqref{pmassl}.

%
%
% the gauge invariance other constraints are also changed in
%
%
%
% \subsubsection{Higher-order version}
%
% \textbf{Bold attempt}
%
% Let us try to generalize the higher-derivative description for spin $s$-field.
% As was explained in \cite{Bekaert:2012vt}, one way to obtain the constraints describing the (off-shell) shadow field
% is to consider the image of the constraints \eqref{nnf}-\eqref{Fgtransf}
% under the automorphism induced by $P\to -\dl{P}$, $\dl{P}\to P$ or $X\to -\dl{X}$, $\dl{X}\to X$.
%
% The image of all the constraints considered before in the above spin-$s$  description, could suggests the following set of constraints (in our case $d$ is even):
% \begin{equation}
% \label{nnft}
%  \Box^{\frac{d-4}{2}+s}\Phi=\bsd\Phi=(U_- -2)\Phi=(N_P-s)\Phi=0\,,
% \end{equation}
% together with the gauge transformation associated to the constraints $\bar\Box\,,\sd\,,\bs\,, \bar T$.
% However, these transformations are compatible with \eqref{nnft} only for gauge parameters obeying similar constraints
% with $s$ replaced with $s-1$.

\subsection{Higher-depth shadow fields}
\label{sec:hd-shad}

The manifestly conformal description of higher-depth shadow fields is obtained by a direct generalization
of the case $t=1$ proposed in~\cite{Bekaert:2012vt}. In ambient terms, one has
\begin{equation}
\label{shad-amb}
\begin{gathered}
\bsd \Phi=0\,, \qquad \big(U_--(t+1)\big)\Phi=0\,,\qquad  \Phi\sim \Phi+(\sd)^{t}\lambda \\
  \Phi\sim \Phi+\bar\Box \chi_1+\bar S\chi_2+\bar T\chi_3\,.
 \end{gathered}
\end{equation}
The precise definition of the system is as follows: one first takes into account the equivalence relation of the second line;
the operators involved in the 2 constraints and in the equivalence relation of the first line are well defined 
on the quotient space defined by the second line. This determines a consistent gauge system. 
Alternatively, one can introduce a BRST operator implementing all the constraints or use the BRST operator only for the constraints {generating gauge transformations} (as explained in~\cite{Bekaert:2012vt}).

To see that the equations \eqref{shad-amb} indeed describe higher-depth shadow fields let us 
expand this system around the point $X^+=1,X^-=x^a=0$. Gauge transformations 
generated by $\bar\Box, \bar S$ and {$\bar T$} are enough to eliminate the dependence 
on $X^-,P^-$ and assume {$T \Phi=0$}. The constraints $U_--t-1$ and $\bsd$  
uniquely fix the dependence on $X^+$ and $P^+$. The remaining gauge 
transformation generated by $(\sd)^t$ is precisely that for depth-$t$ gauge fields. The rigorous 
proof can be given using the parent formulation based on the BRST operator for 
the above system  which also gives a genuine gauge invariant and manifestly 
conformal description of the higher-depth shadow fields. This is spelled out in details in Appendix~\bref{sec:shadows-app}.

Let us also present a concise constrained system describing a reducible multiplet of higher-depth shadow fields.
It is directly related to the one of Section~\bref{sec:reducible} for PM fields. More precisely,
these fields are off-shell boundary values for the respective PM fields. 
As usual, the off-shell boundary values are obtained by replacing all the trace constraints by a gauge equivalence factoring out the same traces, \textit{i.e.} by replacing $\Box,T,S$ by $\bar\Box,\bar S,\bar T$ in the expressions of the constraints and reinterpreting them as gauge generators. In particular, applying this to~\eqref{reducible} we arrive at
\begin{equation}
\begin{gathered}
\label{red-mult-shad}
\bsd \,\Phi=(U_--2)\,\Phi=0\,,\qquad \Phi\sim \Phi+\sd\chi\\
\Phi \sim \Phi+\bar\Box^{\ell_1}\bar S^{\ell_2}\bar S^{\ell_3}\lambda_{\ell_1\ell_2\ell_3}+ (\bar S^2-4\bar\Box\bar T )\lambda_0\,,\quad \ell_1+\ell_2+\ell_3=\ell\,.
\end{gathered}
\end{equation}
It is assumed that one first takes into account the equivalence relation of the second line. As before
the operators in the first line are well-defined in the quotient. The only extra check is to observe that
the equivalence relation associated to ${\bar S}^2-4 \bar \Box \bar T$ doesn't spoil the consistency.
This is clear as ${\bar S}^2-4 \bar \Box \bar T$ commutes with $\sd,\bsd,U_-$.

% 
% 
% The easiest way to see that the system is consistent is first to implement the equivalence relation
% in the second line and observe that 
% operators $X\cdot\d_P$, $P\cdot \d_X$ and $U_-$ act naturally on
% the quotient space off all elements modulo those propositional to 
% ${{\bar \Box}^\ell_1}{\bar S}^{\ell_2}{\bar T}^{\ell_3}$ and
% ${\bar S}^2-4 \bar \Box \bar T$.

\subsection{Partially conserved currents}

In addition to the ambient description of partially conserved currents as current-type boundary values
of PM bulk fields (\textit{i.e.} the constraints~\eqref{pmassl-curr}) there is another one, which generalizes
the ambient description~\cite{Costa:2011mg} of the usual conserved currents: Starting with the constraints~\eqref{pmassl-1} let us apply the automorphism $X\to -\d_X, \d_X\to X$ and exchange the role of some constraints. The result is
\begin{equation}
 T\Phi=(S)^t\Phi=\bsd \Phi=(U_+-3-t)=0\,,\qquad \Phi\sim\Phi+\bar\Box\chi_1+\bar S\chi_2\,.
\end{equation}
This formulation can be seen as an $\mathfrak{o}(d,2)$-covariantization of the equations \eqref{current}
in the sense that the first two constraints are the lift of \eqref{current} and the other four constraints (two imposed, two quotiented) imply that the lift of the current $j$ to the ambient tensor field $\Phi$ is one-to-one.

\subsection{Higher-depth conformal Killing tensors}
\label{sec:Killings}

The complete collection of standard conformal Killing tensors on conformally flat spaces is well known \cite{Geroch} and is generated via the symmetric products of the conformal Killing vectors. Some of their higher-depth generalization{s} are also known. Both are elegantly captured as irreducible modules of the conformal group (see \cite{Eastwood} and refs therein). We recall the latter facts and present a concise proof for them for the sake of completeness.

\begin{lemma}\label{genconfKtens}
In terms of the ambient space $\amb$, all the conformal Killing tensor fields
of spin $s$ and depth $t$
on the flat conformal space of dimension $d>2$ can be obtained via pull-back from
 traceless ambient Killing tensor fields, \textit{i.e.} $\epsilon(X,P)$ satisfying
 \begin{equation}
 \label{Ckill}
  \bsd \epsilon=(\sd)^t \epsilon=(N_X-s+1)\epsilon=(N_P-s+t) \epsilon=\Box \epsilon =S \epsilon=T \epsilon= 0\,.
 \end{equation} 
\end{lemma}
The explicit form reads as
\begin{equation}
\begin{gathered}
(X\cdot\partial_P)\epsilon=(P\cdot\partial_X)^t\epsilon=(X\cdot\partial_X-s+1)\,\epsilon=(P\cdot\partial_P-s+t)\epsilon=0\,,\\
(\partial_P\cdot\partial_X)\epsilon=0\,,\quad (\partial_X\cdot\partial_X)\epsilon=0\,,\quad (\partial_P\cdot\partial_P)\epsilon=0\,.
\end{gathered}
\end{equation} 
It follows the tensor is associated
to a Young tableaux with two rows of length $s-1$ and $s-t$ respectively. Such tensors span the finite-dimensional $\mathfrak{o}(d,2)$-module ${\cal D}(1-s,s-t)$ discussed in Subsection~\bref{sec:group-Killing}.
% In particular for $t=1$ one arrive at rectangular tableax
% asssocated to a usual conformal Killing tensors.
By comparing {to} \eqref{PMGRP} with the extra condition $(N_X-s+1)\lambda=0$, one can see that the above conformal Killing tensors are in one-to-one correspondence with the global reducibility parameters of the respective PM field.

Let us also give a different characterization of the generalized conformal Killing tensors. 
\begin{lemma}
In ambient terms the spin $s$ and depth $t$ conformal Killing tensors can be described by the following constrained system
\begin{equation}
\begin{gathered}
\label{Ckill-dual}
\bsd\epsilon=(\sd)^t\epsilon=(N_X-s+1)\,\epsilon=(N_P-s+t)\epsilon=0\,,\\
\epsilon\sim \epsilon+\bar\Box \chi_1+\bar S\chi_2+\bar T\chi_3\,.
\end{gathered}
\end{equation}
It is understood as follows: 
 %one first passes to the qutotient space determined by the second line and then 
% imposes the constraint of the first line. 
one first implements the equivalence relation of the second line and then
takes the kernel of the operators of the first line in the respective quotient space. 
\end{lemma}
\noindent Note that all the operators present in the first line are well defined in the quotient
as they preserve the ideal of elements generated by $\bar\Box,\bar S,\bar T$.

By comparing with~\eqref{shad-amb} one observes that these are exactly global
reducibility parameters for depth $t$ and spin $s$ shadow fields. Hence, they are 
by definition the respective conformal Killing tensors. The equivalence of the 
above two descriptions is obvious if in~\eqref{Ckill-dual} one can assume that 
$\epsilon$ and parameters $\chi_\alpha$ are polynomial in $X,P$ because in this 
case one can always choose a tracless representative of the equivalence class so 
that~\eqref{Ckill-dual} reduces to~\eqref{Ckill} and hence prove both Lemmas. A 
more careful argument not assuming polynomiality is given in Appendix~\bref{sec:shadows-app}.

% \begin{cor}\label{isomredparpartmass}
% The space of generalized conformal Killing tensor fields of rank $r$ and order $k$
% on the flat conformal spacetime of dimension $d>2$
% is isomorphic to the space of traceless generalized Killing tensor fields of identical rank and order
% on anti de Sitter spacetime of dimension $d+1$.
% 
% This space is described via the irreducible $\mathfrak{o}(d,2)$-module labeled by the
% two-row Young diagram with first row of length $r+k-1$ and second row of length $r$.
% \end{cor}
% 
% In somewhat more physical terms, the space of reducibility parameters for a higher-depth shadow field of spin $s$ and depth $t$ on the flat conformal spacetime of dimension $d>2$
% is isomorphic to the space of reducibility parameters
% for a partially massless field with the same spin and depth
% on anti de Sitter spacetime of dimension $d+1$.
% 
% It is obvious from the above description that generalized conformal Killing tensors are one to one
% with global reducibility parameters for PM fields on $AdS_{d}$. At the same by definition they can be seen
% as global reducibilities for the higher-depth shadow fields in $d$-dimensions. For even $d$ the same applies to higher-depth FT fields.
% 
%The above proposition is a conseuqence of the more general cohomological statements whcih can also be be useful in %differnet contextx.

\section{Background fields for higher-order singletons}
\label{Backgroundfields}

\subsection{Maximal symmetry algebra of higher-order singletons}
 
The corresponding algebra of higher symmetries is already known. 
It was {recently} found by Eastwood-Leistner (for $\ell=2$) \cite{Eastwood:2005} 
and Gover-Silhan (for any integer $\ell$) \cite{Gover:2009} (even the star-product has been studied %in details 
\cite{Michel}) and is the generalization of the celebrated result of Eastwood on the maximal symmetry algebra 
of the d'Alembert equation (for $\ell=1$) \cite{Eastwood:2002su}.
 
The following lemma, which was obtained through the successive works \cite{Eastwood:2002su,Eastwood:2005,Gover:2009}, is important from the present perspective because it should be interpreted
as a new candidate HS algebra. Along the lines
of \cite{Vasiliev:2003ev}, we provide an independent concise proof of this important result in order to be self-contained and to prepare the ground for the analysis of the nonlinear system.

Consider the singleton of order $2\ell$ defined as the conformal primary of weight $\Delta_-=\ell-d/2$
in the kernel of the polywave operator $\Box_0^\ell$. The maximal algebra of its symmetries is the algebra of differential operators ${\cal D}$ on the flat conformal space $\manX_d$ that preserve the polywave equation (in the sense that {$\Box_0^\ell {\cal D} ={\cal C}\Box_0^\ell$ for some differential operator $\cal C$)} modulo the trivial symmetries (defined as ${\cal D}={\cal B}\Box_0^\ell$ for some differential operator $\cal B$).
\begin{lemma}\label{nabhsalg}\cite{Gover:2009}
The maximal algebra of symmetries of the conformal $\ell$-th power of the d'Alembertian
decomposes, as an $\mathfrak{o}(d,2)$-module, as
\begin{equation}
 \algA=\bigoplus\limits_{s=0}^\infty \bigoplus\limits_{m=1}^{\ell} {\cal K}_{s}^{2m-1}
\end{equation} 
where ${\cal K}_{s}^{t}$ is the space of depth $t$ and spin $s$ conformal Killing tensors,%.$K_{s}^{t}$ 
which can be identified with the polynomials satisfying~\eqref{Ckill}, \textit{i.e.} the $\mathfrak{o}(d,2)$-module $\cD(1-s,s-t)$.
% \begin{equation}
%  \begin{gathered}
%  (\d_P\cdot\d_P)\,\epsilon=(\d_P\cdot\d_X)\,\epsilon=(\d_N_X)\,\epsilon=0\\
% (P\cdot\partial_X)\,\epsilon=(P\cdot\d_P-N_X-t+1)\,\epsilon=(P\cdot\d_P-s+1)\,\epsilon=0\,.
% \end{gathered}
% \end{equation} 
% Equivalently, $K_{s}^{t}$ is an irreducible $\mathfrak{o}(d,2)$-modules described by a totally traceless two-row Young diagram where the first line is of length $s-1$ while the second of $s-t$.
\end{lemma}

A direct characterization of the space of higher symmetries is given by the following
\begin{prop}
\label{lemma:char}
As an $\mathfrak{o}(d,2)$-module the algebra of higher symmetries of $\Box_0^\ell$
is isomorphic to the subspace of polynomials in $X^A,P^A$ satisfying
\begin{equation}
\label{reducible-sym}
\begin{gathered}
\bsd \epsilon=U_- \epsilon=\sd  \epsilon=0 \,,\\ (S^2-4T\Box)\ \epsilon=0\,, \qquad 
\Box^{\ell_1}{S}^{\ell_2}{T}^{\ell_3}\epsilon=0\,,
\end{gathered}
\end{equation} 
where $\ell_1+\ell_2+\ell_3=\ell$. %\maxim{minor correction here and below}
\end{prop}
% Note that taking into account the first line the constraints of the second line imply
% \begin{equation}
% \Box^{\ell_1}{S}^{\ell_2}{T}^{\ell_3}\epsilon=0\qquad  
% \end{equation} 
In Appendix~\bref{sec:symm-proof} we give a new proof of Lemma~\bref{nabhsalg}
and the characterization of higher symmetries provided by Proposition~\bref{lemma:char}.
We {use} the first quantized BRST technique where the higher symmetries are identified with inequivalent observables
of the quantized order-$\ell$ singleton described by the BRST operator~\eqref{HOS-BRST-Weyl}. The respective algebra is simply the algebra of its quantum observables.

% The idea of the proof is very similar to the singleton case $\ell=1$ except that the quotient allows to remove the $\ell$th trace only. Details are relagated to Appenix~\bref{sec:symm-proof}

% The group-theoretical description part of the Lemmas \bref{nabhsalg} and \bref{genconfKtens} can be used to show the following isomorphism.
% \begin{cor}\label{isomsymconfKill}
% The space of symmetries of the conformal $\ell$th power of the d'Alembertian is isomorphic to the direct sum
% of the spaces of generalized conformal Killing vector fields of all ranks, and all odd orders from $1$ to $2\ell+1$.
% \end{cor}

Lemma \bref{nabhsalg} suggests the existence of a consistent nonabelian
HS gauge theory based on the Eastwood-Leistner-Gover-Silhan algebra for any integer $\ell$.
The considerations of Subsection~\bref{sec:Killings} then suggests that the corresponding AdS spectrum is a
tower of PM fields of all spins and of odd depths from $1$ to $2\ell-1$.
%Actually, though we chose the conformal signature for definiteness, the Lorentzian signature for the ambient space should also be interesting from a physical perspective. Indeed, de Sitter spacetime might be a better playground because the corresponding HS gauge theory would be unitary (at least this is granted at linearized level). Moreover,
%its holographical dual could be the higher-derivative analogue of the proposal \cite{Anninos:2011ui}.

A further heuristic evidence in favor of the proposal is the $\mathfrak{o}(d,1)$-decomposition match between the 0-form and 1-form modules, 
that should respectively correspond to the ``twisted adjoint'' and ``adjoint'' modules of  the corresponding HS algebra. 

\textbf{Decomposition of the adjoint module:} As explained above, the irreducible $\mathfrak{o}(d,2)$-module 
$\cD(1-s,s-t)$
of on-shell global reducibility parameters of a spin-$s$ and depth-$t$ ($s\geqslant t\geqslant 1$) PM field
is labeled by a two-row Young diagram with first row of length $s-1$ and second row of length $s-t$,
which decomposes as the direct sum of the irreducible $\mathfrak{o}(d,1)$-modules   
labeled by all
two-row Young diagrams with first row of length $\ell_1$ and second row of length $\ell_2$ such that
\begin{equation}
\ell_2\leqslant \ell_1\,,\quad s-t \leqslant \ell_1\leqslant s-1\,,\quad 
0\leqslant \ell_2\leqslant s-t\,.
\end{equation}
For fixed depth $t$ but for spin $s$ ranging over all integers ($s\geqslant t$ are the only non-degenerate cases), one gets the whole collection of all two-row Young diagrams where each of them appears with multiplicity
$t$. Indeed any two-row Young diagram $Y=(\ell_1,\ell_2)$ can be completed to $t$ Young diagrams  $Y_n=(\ell_1+n,\ell_1-t+n)$ where $n=0,1,\ldots, t-1$ which correspond to spins $s=\ell_1+n+1$ ranging from $\ell_1+1$ till $\ell_1+t$.

\textbf{Decomposition of the twisted-adjoint module:} As stated without proof in \cite{Skvortsov:2006at}, the generalized Weyl tensor of a spin-$s$ and depth-$t$ ($s\geqslant t\geqslant 1$) partially massless field spans the irreducible $\mathfrak{o}(d,1)$-module
labeled by a two-row Young diagram with first row of length $s$ and second row of length $s-t$. More generally, the Weyl-module spanned by all independent on-shell gauge-invariant derivatives 
of a spin-$s$ and depth-$t$ partially massless field should be described by the infinite collection of all two-row
Young diagrams with first row of length $\ell_1$ and second row of length $\ell_2$ such that
\begin{equation}
\ell_2\leqslant \ell_1\,,\quad s \leqslant \ell_1\,,\quad s-t+1\leqslant \ell_2\leqslant s\,.
\end{equation}
For fixed depth $t$ but for spin $s$ ranging over all integers, one gets the whole collection of all two-row Young diagrams where each of them appears with multiplicity
$t$. %Indeed, for any $\ell_1$ the length of the second row $\ell_2$ ranges from $s-t+1$ till $s$. 

\subsection{Background fields for a given constrained system}

Given a constrained system (usually a quantized spinning particle or string) there is a natural way to associate to it
a set of gauge fields which, in general,  are subject to nonlinear constraints and nonlinear gauge symmetries. These form a natural set of background fields to which the respective spinning particle or string couples minimally.
This concept was discussed in the context of String Field Theory in~\cite{Horowitz:1986dta} and in~\cite{Segal:2002gd}
for conformal HS fields. Here we mainly follow~\cite{Grigoriev:2006tt} where the case of general constrained system has been considered using BRST framework. In the context of not necessarily conformal HS fields the concept of background fields have been also employed in~\cite{Grigoriev:2006tt,Bekaert:2009ud,Bekaert:2010ky} (see also~\cite{Vasiliev:2005zu}).

Let us consider a quantum constrained system with constraints $T_\alpha(\hat x,\hat p)$. Suppose the constraints are first class so that 
\begin{equation}
 \commut{T_\alpha}{T_\beta}=U^\gamma_{\alpha\beta } T_\gamma\,,
\end{equation} 
for some given operators $U^\gamma_{\alpha \beta}$. Here and below we assume that the operators of spacetime coordinates $\hat x^\mu$ and their conjugate momenta $\hat p_\mu$ are represented in the coordinate representation while the remaining degrees of freedom are represented in terms of the intrinsic representation space $\cH$. 
In this way, the total represenation space is the space of $\cH$-valued wave functions, \textit{i.e.} $\cH$-valued functions of $x^\mu$. 
If $e_A$ denotes a basis in $\cH$ and $e^B$ its dual, then in components we get $T_\alpha=T^A_{\alpha\, B}(\hat x,\hat p) e_A\tensor e^B$. Instead of working with differential operators in $x^\mu$ we choose to work in terms of their Weyl symbols depending on $x,p$ and hence replace operator multiplication with the Weyl star product (tensored with the operator product in the internal space).

\newcommand{\Ttheta}{\Theta}
Given this data consider generating functions $\Ttheta_\alpha(x,p)=\Ttheta^A_{\alpha B}(x,p)e_A\tensor e^B$ for background fields and subject them to the following equations:
\begin{equation}
\label{b-eom}
 \qcommut{\Ttheta_\alpha}{\Ttheta_\beta}-\Upsilon_{\alpha\beta}^\gamma(x,p) * \Ttheta_\gamma=0
\end{equation}
along with the gauge symmetries
\begin{equation}
\label{b-gs}
 \delta \Ttheta_\alpha=\qcommut{\Ttheta_\alpha}{\lambda}+\chi^\gamma_\alpha *\Ttheta_\gamma\,.
\end{equation}
Here, $\Upsilon_{\alpha\beta}^\gamma$ are symbols of some operators and $\lambda(x,p),\chi(x,p)$ are gauge parameters. Note that~\eqref{b-gs} are natural gauge symmetries of~\eqref{b-eom}.
%Let us comment on the role played by $U$. In contrast to generating functions $\Ttheta^A_B$

The above relation and symmetry are nothing but the well-known defining relation for the first class constraints and the natural equivalence in the choice of constraints. Now these are interpreted as equations of motion and gauge symmetries of a field theory whose fields are encoded in $\Ttheta_\alpha$ and whose gauge parameters are in $\lambda,\chi^\alpha_\beta$. 
These fields have a natural interpretation as background fields coupling minimally to the quantized ``particle'' described by $T_\alpha$.
%Alternatively, one can think of them as background field for the free gauge system determined in the %first-quantized formalism by the same quantum constrained system.

% 
% 
% 
% 
% 
% 
% Note that this theory is background independent in the sense that equations and gauge transformations do not involve anyextra background fields. In this sense  this theory itself is a theory of background fields.
To make the relation of the above background fields to the starting point quantum constrained system
more explicit, one observes that $\Ttheta_\alpha=T_\alpha$ is a particular solution to~\eqref{b-eom} for which
$\Upsilon_{\alpha\beta}^\gamma=U_{\alpha\beta}^\gamma$. Interpreting it as a {vacuum} solution it is instructive to linearize the system around.  The linearized equations of motion take the form (here $\Psi_\alpha$ denotes the perturbation, \textit{i.e.} 
$\Ttheta_\alpha=T_\alpha+\Psi_\alpha$)
\begin{equation}
\label{lin-eom}
 \qcommut{T_{[\alpha}}{\Psi_{\beta]}}-U_{\alpha\beta}^\gamma *\Psi_\gamma-u_{\alpha\beta}^\gamma *T_\gamma=0\,,
\end{equation}
for some operators $u^\gamma_{\alpha\beta}$. For the linearized gauge tranformations one gets
\begin{equation}
\label{bf-gs}
 \delta\Psi_\alpha= \qcommut{T_\alpha}{\lambda}+\chi^\gamma_\alpha * T_\gamma \,.
\end{equation}

Let us note that it is often possible to consistently truncate the gauge symmetries of the background fields
by \textit{e.g.} putting to zero the parameters $\chi^\alpha_\beta$ in \eqref{bf-gs}. For instance the nonlinear off-shell system
proposed in~\cite{Vasiliev:2005zu} to describe off-shell higher spin fields on Minkowski space belongs to this class (see~\cite{Grigoriev:2006tt} for details on the relationship). Let us mention also the nonlinear off-shell systems~\cite{Grigoriev:2011gp} for (AdS) higher spin fields which also belong to this class.

% 
% The consistent nonlinear systems~\cite{Vasi} of background fields of this type are relevant for Fronsdal-type (in conrast to conformal) higher spin gauge fields at the interacting level.
% HS have been proposed in~\cite{Vasi,Gr} to describe 

% 
% Essentially this trancation
% underlies the system 

\subsection{Global symmetries and deformations}
The linearized equations~\eqref{lin-eom} and \eqref{bf-gs} can be given an alternative interpretation
solely in terms of the starting point constrained system. Indeed, given a constrained system with the constraints $T_\alpha$ one naturally considers global symmetries of the equations of motion $T_\alpha(\hat x,\hat p) \Phi=0$. These are given by symbols $\lambda$ satisfying
\begin{equation}
\label{grp-bf}
\qcommut{T_\alpha}{\lambda}+\chi^\gamma_\alpha * T_\gamma =0\,, \qquad \lambda \sim \lambda +\rho^\alpha * T_\alpha
\end{equation}
for some $\chi_\alpha^\gamma, \rho^\alpha$. 

% It follows that the global symmetries are one to one with the global reducibility identitis for the gauge transformation of the background fields. Indeed, requiring
% $\delta\Psi_\alpha=0$ in \eqref{bf-gs} one gets the above condition. The above equivalence relation is a natural
% equivalence of gauge parameters.

At the same time taking $\delta\Psi=0$ in~\eqref{bf-gs} one arrives at the equations for global reducibility parameters
$
%\begin{equation}
 \qcommut{T_\alpha}{\lambda}+\chi^\gamma_\alpha*T_\gamma =0\,.
%\end{equation}
$
Taking into account gauge equivalence of reducibility parameters it follows the global symmetries are in one-to-one correspondence
with the global reducibility parameters for the {linearized} gauge transformation of the background fields.  Furthermore,
relations~\eqref{grp-bf} are nothing but a definition of inequivalent observables of the quantum constrained system.
To conclude, the following three spaces coincide: observables of the quantum constrained system,
reducibility relations for the associated background fields, and global symmetries of the equations of motion $T_\alpha \Phi=0$. As we are going to see all these three spaces are different interpretations of a single cohomology group.
% This equation is identical to the condition that $\lambda$ is an observable of the starting point constrained system.
% Takin into account natural equivalence of the gauge parameters one concludes that global symmetries of the equations of motion $T_\alpha\Phi=0$ are one to one with the first quantized observalbles.
% 

For the constrained system with constraints $T_\alpha$, one can study its consistent deformations, \textit{i.e.}
infinitesimal deformations of the consraints $T_\alpha$ such that they still form a closed algebra and
hence still determine a consistent constrained system. Requiring $T_\alpha+\Psi_\alpha$ to form a close algebra
gives:
\begin{equation}
\qcommut{T_{[\alpha}}{\Psi_{\beta]}}+U_{\alpha\beta}^\gamma * \Psi_\gamma+ u_{\alpha\beta}^\gamma * T_\gamma=0
\end{equation}
where we have explicitly introduced first order deformation $u$ of $U$. Comparing with \eqref{b-eom}
and taking into account gauge equivalence one finds that first order deformations of the constrained system
are one-to-one with the configurations of the associated background fields.

\subsection{Batalin--Fradkin--Vilkovisky description of background fields}

The description of background fields can be performed in a very concise and mathematically clear form using the BFV-BRST language. Using ghost variables $c^\alpha$ and their conjugate momenta $b_\alpha$ one introduces
nilpotent BRST operator
\begin{equation}
 \Omega=c^\alpha T_\alpha-\half c^\alpha c^\beta U_{\alpha \beta}^\gamma b_\gamma+\ldots
\end{equation}
where dots denote terms of order 2 and higher in $b_\alpha$. As before we work in terms of operator symbols. In particular for ghosts $c,b$ we choose the ``normal'' ordering (more precisely, corresponding to $c$ on the left and $b$ on the right).

Equations \eqref{b-eom} and \eqref{b-gs} appear as component equations of respectively
\begin{equation}
 \qcommut{\Xi}{\Xi}=0\,, \quad \delta\Xi=\qcommut{\Xi}{\Lambda}\,,\qquad \gh{\Xi}=1\,,\quad \gh{\Lambda}=0\,.
\end{equation}
Here $\Xi$ and $\Lambda$ are general symbols of the above ghost degrees. The identification of components is as follows
\begin{equation}
 \Xi=c^\alpha\Ttheta_\alpha-\half c^\alpha c^\beta \Upsilon_{\alpha\beta}^\gamma b_\gamma+\ldots\,,\qquad
 \Lambda=\lambda+c^\alpha \chi_\alpha^\beta b_\beta+\ldots
\end{equation}
where dots denote terms of higher powers in ghosts. The system linearized around a background (vacuum) solution $\Xi=\Omega$
reads as
\begin{equation}
 \qcommut{\Omega}{\Psi}=0\,, \qquad \delta \Psi=\qcommut{\Omega}{\Lambda}\,,
\end{equation} 
where $\Psi=c^\alpha \Psi^\alpha-\half c^\alpha c^\beta u_{\alpha\beta}^\gamma b_\gamma+\ldots$.
These respectively contain equations~\eqref{lin-eom} and \eqref{bf-gs} as component equations at lowest orders in ghosts.

In contrast to the discussion in the previous sections where only $\Ttheta_\alpha$ were considered as the generating functions for background fields now all component fields entering $\Xi$ (or $\Psi$ for the linearized system) are at the same footing so that at first glance we have  more fields. A similar remark applies to gauge parameters. However, all the higher components in $\Xi$ and $\Lambda$ turn out to be generalized auxiliary. Indeed, if $\Omega$ is proper (\textit{i.e.} all reducibility relations are taken into account) they are uniquely determined by the lowest components $\Psi_\alpha$ modulo gauge equivalence. This is also the case if one considers a full nonlinear system around the vacuum $\Omega$. Indeed, it is a standard statement in the theory of constrained systems that the BRST operator is uniquely determined by the constraints modulo natural gauge equivalence so that all the (higher) structure functions of the gauge algebra along with the higher components in $\Lambda$ are auxiliary fields, 
St\"
uckelberg fields and associated gauge parameters from the present perspective. 

Assuming $\Omega$ proper, one concludes that inequivalent configurations for background fields are one-to-one with the cohomology of the adjoint action of $\Omega$ at ghost degree $1$. In this form it is obvious that these are the same as consistent deformations.

In BRST terms, global reducibility parameters of the linearized background fields are described by
\begin{equation}
 \qcommut{\Omega}{\Lambda}=0\,, \quad \Lambda \sim \Lambda+\qcommut{\Omega}{\Xi}\,,\qquad \gh{\Lambda}=0\,,\quad \gh{\Xi}=-1\,,
\end{equation}
and hence coincide with adjoint cohomology at ghost degree $0$. These are global symmetries of the equations encoded
in $\Omega\Phi=0$ with $\gh{\Phi}=0$. Alternatively, from the first quantized point of view these are inequivalent observables.

In the same way inequivalent configurations of the linearized background fields are described by
\begin{equation}
 \qcommut{\Omega}{\Psi}=0\,, \quad \Psi \sim \Psi+\qcommut{\Omega}{\Lambda}\,,\qquad \gh{\Psi}=1\,,\quad \gh{\Lambda}=0\,,
\end{equation} 
and hence are inequivalent consistent deformations of the constrained system described by $\Omega$.

The following important comment is in order: with a given operator $\Omega$ one can associate a family of gauge invariant equations of motion. Indeed, consider equations of motion and gauge transofrmations
\begin{equation}
\label{omega-phi}
 \Omega \Phi(x,c)=0\,,\quad \Phi \sim \Phi+\Omega \epsilon(x,c)\,, \qquad \gh{\Phi}=k\,,\quad \gh{\epsilon}=k-1
\end{equation}
where $k$ is an integer and for definiteness we have represented ghost variabels in the coordinate representation so that $\hat c^\alpha=c^\alpha$ and $\hat b_\alpha =\dl{c^\alpha}$. The same result can be achieved by choosing momenta representation for some of the ghost variables. In terms of the starting point constraints this means taking one or another subset of $T_\alpha$ to generate gauge transformations.

Although, gauge invariant equations encoded by $\Omega$ depend on $k$ this is 
not the case for background fields, consistent deformations, and global 
symmetries. Indeed, these are determined by the adjoint action of $\Omega$ in 
the algebra of symbols and hence do not depend on the choice of representation. 
One then concludes that all these objects are actually associated to a family of 
gauge invariant equations.

The background fields encoded in $\Xi$ can also be seen as background fields for free gauge field $\Phi$
determined by~\eqref{omega-phi}. Indeed the equations of motion and gauge symmetries for $\Phi$ 
over the background $\Xi$ are obtained by replacing $\Omega$ with $\Xi$ in~\eqref{omega-phi}, where e.g. $\Xi\Phi$
is understood as the action of operator associated to symbol $\Xi$ on state $\Phi$. Note that
global symmetries of gauge fields~\eqref{omega-phi} are precisely the symmetries of the vacuum solution
$\Xi=\Omega$. If one restricts to hermitean $\Xi$ the coupling of $\Phi$ to background fields encoded in $\Xi$ is simply described by the action $S=\half\inner{\Phi}{\Xi \Phi}$. In the later case the ghost degree of $\Phi$
should be compatible with that of the inner product.

Let us finally comment on the unfolded formulation for the background fields. 
According to~\cite{\BGST,\BGadS} the fields of the unfolded formulation is the 
cohomology of the starting point BRST operator in the formal space where $x$ 
variables have been replaced with the formal $y$-variables. More precisley, 
basis elements of the BRST cohomology at ghost degree $p$ give rise to $p$-form fields.
In particular, the sector of $0$-forms originates from inequivalent solutions to equations of motion in the space of formal power series in $y$. For the BRST operator $D=\qcommut{\Omega}{\cdot}$ determining the linearized theory of background fields these are describe by cohomology at ghost degree $1$ because we use the convention $\gh{\Psi}=1$
(in the more conventional case fields are found in $0$ degree component).
Furthermore, 1-form sector (gauge module) is associated to cohomology at ghost degree $0$
and hence are one-to-one with observables of the starting point quantum constrained system (which are, in turn,
one-to-one with global symmetries of the associated linear system $\Omega\Phi=0$). Of course, the sector of 1-forms
is also one-to-one with global reducibility relations for linearized background fields.

\subsection{Background fields for the higher-order singleton}
We now consider a concrete quantum constrained system, namely the higher-order singleton.  The Weyl symbol of the respective BRST operator is given in~\eqref{HOS-BRST-Weyl}. Its adjoint action has the following structure:
%\begin{multline}
\begin{equation}
\label{D-adj}
 D=\qcommut{\Omega}{\cdot}=c_1 \qcommut{{\bar T}^\ell}{\cdot}+c_0 U_-+c_{-1} \bsd
 +{\bar T}^\ell \dl{b_1}+\bar S \dl{b_0}+\bar\Box\dl{b_{-1}}+\ldots
\end{equation}
%\end{multline}
It is clear that for $\ell=1$ this is precisely the BRST operator describing a multiplet of shadow fields (strictly speaking there are terms of higher order in derivatives but they do not change the result). 

To see what does this system describe for $\ell>1$ let us analyze the cohomology at ghost degree $0$. This is precisely the higher symmetries of order $\ell$ singleton. This suggests that $D$ describes a reducible multiplet of higher-depth shadow fields given by~\eqref{red-mult-shad}. Indeed, the global reduciblity parameters
of this multiplet precisely match the spectrum of higher symmetries. To put it differently, these background fields are boundary values of the reducible multiplet of the bulk PM fields described in Subsection~\bref{sec:reducible}.
We do not have an exhaustive proof of this statement for generic $\ell$ but the given evidences strongly 
support this proposal. 

Note that according to the general discussion of background fields \eqref{D-adj} describes the linearization of a nonlinear off-shell system of shadow fields whose field content matches the multiplet of the higher symmetry algebra. One should also expect that in $d=2m$ these fields admit nonlinear gauge invariant equations based on the higher symmetry algebra. 

% 
% 
% 
% This statement also provides additional argument that the space of configurations
% (\textit{i.e.} cohomology at degree $1$) also matches that for the multiplet of  
% 
% 
% 
% 
% Using the appropriate degree one can assume that the first three terms in the second line (strictly speaking along with the higer derivative extra terms) for a lowest degree ($-1$) term in $\Omega$. By reduing to its cohomology one can assume the representtive to be independent of ghost momenta $b$, $P^-$, $\ell$-th and higher powers of $Y^-$,
% and to be annihilated by $T$. A homogeneous in $Y^-$ representative of this type has the form $\Phi={Y^-}^t\phi$
% where $T\phi=0$ and $\phi$ is $b,P^-$-independent. One then reduces to the cohomology of the term in the BRST operator containing the second and the third constraints of te first line. This simply amounts to taking the kernmel of these two operators (up to ghost modifications). One then shows that any such $\Phi$ can be completed to $\Phi^\prime$
% by terms depending on $P\cdot Y$ and $P^2$ such that $\bsd \Phi^\prime=0$. $\Phi^\prime$ by construction belongs to
% the space of elements whose total degree in $Y^-$, $P^-$ and the depth of $T$ trace (\textit{i.e.} the minimal $m$ such that $T^m\Phi^\prime=0$) is not greater then $\ell-1$.
% 
% This suggests that the constrained system encoded in $\Omega$ is equivalent to the following one:
% \begin{equation}
%  U_-,\quad \bsd\,,\quad \sd\,,\quad {\bar T }^{\ell_1}{\bar\Box}^{\ell_3}{\bar S}^{\ell 2}
% \end{equation}
% 

\section{Conclusion}

As a conclusion, let us summarize our main results and then present some {speculations based on them.}

A gauge and $\mathfrak{o}(d,2)$ covariant relation between bulk and boundary values of partially massless fields has been developed. The approach was based on identifying AdS or conformal gauge fields as those associated to a suitable  first-quantized constrained system defined in the ambient space. More technically, we employed the first-quantized BFV-BRST method along with its parent extension and homological tools. 

The leading boundary behavior of a partially massless field determines a higher-depth shadow field on the boundary while the subleading data determines a partially conserved current. When the dimension of the boundary is even, an obstruction appears to the extension of the asymptotic solution in the bulk via power series of the radial variable. This obstruction can be removed if some generalized Fradkin--Tseytlin equations, suitable for higher-depth conformal gravity, are imposed on the shadow fields. This situation is the precise analogue of the standard situation for scalar singletons (of any order). As a byproduct, manifestly conformal descriptions of higher-order singletons, partially conserved currents, higher-depth shadow fields and Fradkin--Tseytlin equations, have been presented.

In the special case of $AdS_5$ we observe that the boundary values of low spin $s=1,2$ and maximal depth ($t=s$)
partially massless fields are shadow fields subject to the higher-depth Fradkin--Tseytlin equations which in this case are second order and are those
identified by Deser and Nepomechie~\cite{Deser:1983mm}. These can again be sees as PM fields of maximal depth in $AdS_4$. {This correspondence does not seem to extend to higher spins though.}

Via purely group-theoretical techniques, the tensor product of two singletons of order $\ell$ has been shown to decompose into the direct sum of partially conserved currents
of all ranks and of odd depth from 1 till $2\ell-1$. This decomposition is in agreement with the higher symmetries of the polywave equation \eqref{polywave} which have been shown to be in one-to-one correspondence with the global symmetries of the corresponding partially massless fields and higher-depth shadow fields.
This algebra of higher-order singleton symmetries defines a candidate higher-spin algebra, which is a quotient of the off-shell bosonic HS algebra. Therefore the Vasiliev equations of bosonic HS gravity in any dimension based on this algebra should remain consistent by construction.
For instance, the decomposition of the adjoint and twisted-adjoint representations appear to match in the usual way.

All these results support the following generalization of the conjectures in \cite{Sezgin:2002rt}: the large-$N$ limit of
the $O(N)$ vector model in $d$ dimensions at a multicritical isotropic Lifshitz point\footnote{A seminal study of the large-$N$ limit of Lifshitz points was provided in the paper \cite{Hornreich} (see also \cite{Shpot:2012mw}
and references therein to recent works).} of order $2\ell$ (in the derivative expansion) might be dual to a HS theory around $AdS_{d+1}$
with an infinite tower of partially massless symmetric tensor fields of spins 0, 2, 4, ... and of depths 1, 3, ..., $2\ell-1$.
More precisely, the $N$ fundamental scalar fields fit in a $N$-vector $\vec\phi$ whose scaling dimension is $\Delta_-=\frac{d}2-\ell$ in dimension $d$,
which is nothing but the engineering dimension consistent with the higher-derivative kinetic term $\int \vec\phi\cdot(\Box_0)^\ell\vec\phi\,d^dx\,$.
The composite field driving the double trace deformation $(\vec\phi^2)^2$ is the ``spin-$0$ current'' $\vec\phi^2$ whose scaling dimension is equal to: its engineering dimension $\Delta_{\mbox{free}}=2\Delta_-=d-2\ell$ at the multicritical Gaussian fixed point, its conjugate dimension $\Delta_{\mbox{int}}=d-\Delta_{\mbox{free}}=2\ell$
at the multicritical isotropic Lifshitz point in the large-$N$ limit.\footnote{The scaling dimension at the interacting fixed point is anomalous ($\Delta_{\mbox{int}}\neq \Delta_{\mbox{free}}$) except when $d=4\ell$. The latter exception should indicate the ``triviality'' (merging of the Gaussian and interacting fixed points) of the theory in $d=4\ell$ dimensions. Notice that for $d<4\ell$ one gets $\Delta_{\mbox{int}}> \Delta_{\mbox{free}}$ thus, inside the multicritical surface, the Gaussian and interacting fixed points are respectively repulsive and attractive in the infrared.%, so that one expects to have a renormalization group trajectory flowing from one to the other.
} The unbroken HS theory, for the $\Delta=\Delta_{\mbox{free}}$ boundary condition on the bulk scalar field, described by the Vasiliev equations \cite{Vasiliev:2003ev} based on the higher-order singleton symmetry algebra should be dual to the multicritical Gaussian fixed point.
The $\Delta=\Delta_{\mbox{int}}$ boundary condition on the bulk scalar field breaks the HS symmetries at finite $N$ and should correspond to the multicritical interacting fixed point.

The dS/CFT version of the above proposal should be also of interest. Remember that, though partially massless fields are not unitary on AdS, their dS analogues are. Following the conjecture of \cite{Anninos:2011ui}, one might further speculate that the Euclidean $Sp(N)$ vector model with anticommuting scalars at multicritical isotropic Lifshitz points might be dual to (unitary) HS theories of partially massless tensor fields around de Sitter spacetime.

It is worth mentioning that the constructions in the present paper should generalize smoothly to general mixed symmetry fields on AdS using the formulation developped in~\cite{Alkalaev:2009vm,Alkalaev:2011zv}. Indeed, all the basic structures are exactly the same so that the generalization is of purely technical nature. In particular, it is natural to expect that the match between AdS gauge fields and their (off-shell or on-shell) boundary values
extends to the mixed symmetry case and moreover provides a manifestly $\mathfrak{o}(d,2)$ and gauge covariant description for generic conformal gauge fields.

%\newpage

\section*{Acknowledgments}

We are grateful to G.~Barnich, E.~Joung, O.~Shaynkman, E.~Skvortsov, I.~Tipunin for discussions. 
X.~B. is thankful to C.~Bervillier, N.~Boulanger and C.~Lecouvey while 
M.~G. thanks K.~Alkalaev, A. Chekmenev, R.~Metsaev, K.~Mkrtchyan and M.~A.~Vasiliev.

X.B. gratefully acknowledges the Dynasty Foundation (Russia) for the support of his visit to
the Lebedev Institute (Moscow) where this work was initiated. We thank the Galileo Galilei Institute for Theoretical Physics (Florence) for the hospitality during
the completion of this work and acknowledge the partial support from INFN (Italy) and the Dynasty Foundation (Russia).
The work of M.G. was supported by RFBR grant 11-01-00830 . 
\appendix

\section{Proof of the generalized Flato--Fronsdal theorem}
\label{sec:FFproof}

In order to prove the generalization of the Flato--Fronsdal theorem for higher-order singletons,
some further representation technology is convenient.
The idea of the proof is to show that the characters of both sides of the equation in Proposition \bref{prop:FF}
are equal. This strategy was also adopted in the original paper of Flato and Fronsdal \cite{Flato:1978qz} and in
its extensive generalization \cite{Dolan:2005wy}
for various unitary $\mathfrak{o}(d,2)$-modules.
Actually,
the explicit formulae for the character
$\chi\left(\Delta,Y\right)$ of the module ${\cal D}\left(\Delta,Y\right)$
will only be provided in the simplest case $d=3$ as concrete illustrations of our proof.
Technically speaking, we will circumvent the explicit computation of characters for $d>3$ by making use of
the decomposition into finite-dimensional $\mathfrak{o}(2)\oplus\mathfrak{o}(d)$-modules.
This strategy is equivalent to the computation of characters because the decomposition of the Verma modules considered in Proposition \bref{prop:FF} only involves finite multiplicities
(thus this decomposition is exactly equivalent to the character and it therefore characterizes uniquely the $\mathfrak{o}(d,2)$-module). Such a strategy was followed in the early generalization \cite{Heidenreich:1980xi} of the Flato--Fronsdal theorem to the case $d=4$.

The notation ${\cal Y}\left(\Delta,Y\right)$ for
irreducible finite-dimensional $\mathfrak{o}(2)\oplus\mathfrak{o}(d)$-modules can be extended to
(possibly) reducible finite-dimensional modules with the following rules for the direct sum and the tensor product:
\begin{eqnarray}
{\cal Y}\left(\Delta,Y_1\right)&\oplus& {\cal Y}\left(\Delta,Y_2\right)\,\, =\,\,{\cal Y}\left(\Delta,Y_1\oplus Y_2\right)\nonumber\\
{\cal Y}\left(\Delta_1,Y_1\right)&\otimes& {\cal Y}\left(\Delta_2,Y_2\right) \,=\,{\cal Y}\left(\Delta_1+\Delta_2,Y_1\otimes Y_2\right)
\label{sumprod}
\end{eqnarray}
For instance, these rules can be used to show the

\begin{lemma}\label{decompVerma}
The decomposition of the Verma $\mathfrak{o}(d,2)$-module  ${\cal V}\left(\Delta,Y\right)$ into
an (infinite) sum of finite-dimensional $\mathfrak{o}(2)\oplus\mathfrak{o}(d)$-modules is
\begin{equation}\label{Vmoduldecomp}
{\cal V}\left(\Delta,Y\right)=\bigoplus_{p,q=0}^\infty {\cal Y}\left(\Delta+p+2q,p\otimes Y\right) \,.
\end{equation}
\end{lemma}

\proof{The raising ladder operators $\textsc{J}^+_i$ in the definition \eqref{Vmoddef} of the Verma $\mathfrak{o}(d,2)$-module ${\cal V}\left(\Delta,Y\right)$ raise by one unit the energy $\mathfrak{o}(2)$ and carry the fundamental representation of $\mathfrak{o}(d)$.
In other words, the algebra ${\mathbb R}^{d}=$span$\{\textsc{J}^+_i\}$ of raising ladder operators
is as a finite-dimensional $\mathfrak{o}(d,2)$-module isomorphic to ${\cal Y}\left(1,1\right)$.
The Poincar\'e-Birkhoff-Witt basis of the universal enveloping algebra ${\cal U}\Big({\mathbb R}^{d}\Big)$ are the (symmetric) powers $\textsc{J}^+_{i_1}\ldots \textsc{J}^+_{i_r}$ of the (commuting) ladder operators.
As an $\mathfrak{o}(d,2)$-module, the algebra ${\cal U}\Big({\mathbb R}^{d}\Big)$ on ${\cal V}\left(\Delta,Y\right)$ can be decomposed as the
(infinite) sum of finite-dimensional $\mathfrak{o}(2)\oplus\mathfrak{o}(d)$-modules
$$
{\cal U}\Big({\mathbb R}^{d}\Big)\cong \bigoplus_{p,q=0}^\infty {\cal Y}\left(p+2q,p\right)
$$
where $r=p+2q$ corresponds the total number of ladder operators, $p$ to the number of free indices and $q$ corresponds to the number of contractions of indices via the $\mathfrak{o}(d)$-metric $\delta^{ij}$.
In the definition \eqref{Vmoddef} of the Verma module ${\cal V}\left(\Delta,Y\right)$, the action of ${\cal U}\Big({\mathbb R}^{d}\Big)$ on ${\cal V}\left(\Delta,Y\right)$ is identical to a tensor multiplication, thus
$$
{\cal V}\left(\Delta,Y\right)\cong \bigoplus_{p,q=0}^\infty {\cal Y}\left(p+2q,p\right)\otimes{\cal Y}\left(\Delta,Y\right)
$$
The lemma is obtained by applying the rule \eqref{sumprod}.

In CFT language, the index $p$ would correspond to the number of traceless partial derivatives of the conformal primary
corresponding to ${\cal Y}\left(\Delta,Y\right)$ while the index $q$ corresponds
to the power of the wave operator.
}

\noindent\textbf{Example:} In order to present some explicit
character formulae in the particular case $d=3$, let us introduce
the variables $\alpha=\exp(i\,\textsc{E}/2)$ and
$\beta=\exp(i\,\textsc{L}_{12}/2)$, following the notations of
\cite{Flato:1978qz}. From lemma \bref{decompVerma}, one computes the
character of the Verma $\mathfrak{o}(3,2)$-module and finds the
rational function
\begin{eqnarray}\label{charVerma}
 \chi\left({\cal V}\left(\Delta,s\right)\right)
&=&\bigoplus_{p,q=0}^\infty\alpha^{2(\Delta+p+2q)}\chi_p (\beta)\chi_s (\beta)\nonumber\\
&=&-\,\frac{\alpha^{2\Delta-3}\chi_s (\beta)}{(\alpha-\alpha^{-1})\big(\alpha\beta-(\alpha\beta)^{-1}\big)\big(\frac{\alpha}{\beta}-\frac{\beta}{\alpha}\big)}
\end{eqnarray}
where the character of the spin-$s$ irreducible $\mathfrak{o}(3)$-module is given by
\begin{equation}
\chi_s(\beta)=\frac{\beta^{2s+1}-\beta^{-(2s+1)}}{\beta-\beta^{-1}}\,.
\end{equation}

The tensor product of finite-dimensional irreducible $\mathfrak{o}(d)$-modules can be evaluated through the Littlewood-Richardson and branching
rules\footnote{For a concise and self-contained review on the Young diagrammatic technology for computing products and decompositions of irreducible representations of the general linear and orthogonal groups, see \textit{e.g.} Section 4 of \cite{Bekaert:2006py}.} in order to decompose the quotients ${\cal D}\left(\Delta,Y\right)$ as a sum of
irreducible  $\mathfrak{o}(2)\oplus\mathfrak{o}(d)$-modules.
The following two lemmas perform this decomposition for the (higher-order) singletons and the (partially) conserved conformal currents.

\begin{lemma}\label{confscal}
The irreducible $\mathfrak{o}(d,2)$-module ${\cal D}\left(\frac{d}2-\ell,0\right)$ of the higher-order scalar singleton of order $2\ell$
decomposes as the infinite sum of
finite-dimensional $\mathfrak{o}(2)\oplus\mathfrak{o}(d)$-modules
$${\cal D}\left(\frac{d}2-\ell,0\right)=\bigoplus_{p=0}^\infty\bigoplus_{q=0}^{\ell-1} {\cal Y}\left(\frac{d}{2}-\ell+p+2q,p\right)$$
\end{lemma}

\proof{The respective decomposition of the Verma modules %${\cal V}\left(\frac{d}2\pm \ell,0\right)$
$${\cal V}\left(\frac{d}2\pm \ell,0\right)=\bigoplus_{p,q=0}^\infty {\cal Y}\left(\frac{d}2\pm \ell+p+2q,p\right)$$
follows from the formula \eqref{Vmoduldecomp}. The decomposition of the higher-order singleton \eqref{highderscalquot}
is obtained via the substraction of all the $\mathfrak{o}(2)\oplus\mathfrak{o}(d)$-modules
in the Verma module
${\cal V}\left(\frac{d}2- \ell,0\right)$
that also appear in the Verma module ${\cal V}\left(\frac{d}2+ \ell,0\right)$.
This is readily seen by the equalities
\begin{eqnarray}
&&{\cal V}\left(\frac{d}2- \ell,0\right)=\bigoplus_{p,q=0}^\infty {\cal Y}\left(\frac{d}2- \ell+p+2q,p\right)\nonumber\\
&&=\bigoplus_{p=0}^\infty\left(\,\bigoplus_{q=0}^{\ell-1} {\cal Y}\left(\frac{d}{2}-\ell+p+2q,p\right)\,\oplus\,\bigoplus_{q=\ell}^{\infty} {\cal Y}\left(\frac{d}2- \ell+p+2q,p\right)\,\right)\nonumber\\
&&=\bigoplus_{p=0}^\infty\left(\,\bigoplus_{q=0}^{\ell-1} {\cal Y}\left(\frac{d}{2}-\ell+p+2q,p\right)\,\oplus\,\bigoplus_{q'=0}^{\infty} {\cal Y}\left(\frac{d}2+ \ell+p+2q',p\right)\,\right)\nonumber\\
&&=\left(\,\bigoplus_{p=0}^\infty\bigoplus_{q=0}^{\ell-1} {\cal Y}\left(\frac{d}{2}-\ell+p+2q,p\right)\,\right)\,\oplus\,
{\cal V}\left(\frac{d}2- \ell,0\right)\nonumber
\end{eqnarray}
}

\noindent\textbf{Example:} From Lemma \bref{confscal}, one computes
that the irreducible $\mathfrak{o}(3,2)$-module ${\cal
D}\left(\frac32-\ell,0\right)$ of the $d=3$ singleton of
order $2\ell$ has character given by the rational function
\begin{align}\label{charformcs}
 &\chi\left({\cal D}\left(\frac32-\ell,0\right)\right)\,=\,\sum\limits_{p=0}^\infty\sum\limits_{q=0}^{\ell-1} \alpha^{2(\frac32-\ell+p+2q)} \chi_p (\beta)
\,=\,
\\ \nonumber
&\frac{(\alpha^{2\ell}-\alpha^{-2\ell})}{(\alpha-\alpha^{-1})\big(\alpha\beta-(\alpha\beta)^{-1}\big)\big(\frac{\alpha}{\beta}-\frac{\beta}{\alpha}\big)}=
\chi\left({\cal Y}\left(\frac32-\ell,0\right)\right)-\chi\left({\cal Y}\left(\frac32+\ell,0\right)\right)
\end{align}
where the last equality is the character translation of \eqref{highderscalquot} and follows from \eqref{charVerma}.

The tensor product of finite-dimensional irreducible $\mathfrak{o}(d)$-modules can be evaluated through the Littlewood-Richardson rule \textit{together with} the branching rule from $\mathfrak{gl}(d)$ to $\mathfrak{o}(d)$ modules. In order to formulate the next lemma, one needs the following notation: $Y_1\otimes^\prime Y_2$ stands for the
sum of Young diagrams obtained by applying the Littlewood-Richardson rule only.

\begin{lemma}\label{partconscur}
The irreducible $\mathfrak{o}(d,2)$-module
${\cal D}\left(d+s-t-1,s\right)$ of the partially conserved current
of rank $s$ and depth $t$
decomposes as the infinite sum of
finite-dimensional $\mathfrak{o}(2)\oplus\mathfrak{o}(d)$-modules
$${\cal D}\left(d+s-t-1,s\right)=\bigoplus_{l,m=0}^\infty\,\bigoplus_{n=\max\{0,s-t+1\}}^{s} {\cal Y}\big(d+2(s+l)-t-1+m-n,m\otimes^\prime n\big)$$
\end{lemma}

\proof{The decomposition of the partially conserved current \eqref{partconscurrmod} follows the same procedure but
the substraction of the $\mathfrak{o}(2)\oplus\mathfrak{o}(d)$-modules appearing in ${\cal I}\left(d+s-1,s-t\right)$ from the decomposition of ${\cal V}\left(d+s-t-1,s\right)$ is more involved
because they correspond to contractions in the tensor product of Young diagrams (\textit{i.e.} divergences in CFT language).

The following fact should be used in the proof: the decomposition of the tensor product $r_1\otimes r_2$ of two $\mathfrak{o}(d)$-modules labeled by single rows of respective lengths $r_1\geqslant r_2$ reduces to the application of the Littlewood-Richardson rule for each term in the sum
$$r_1\otimes r_2\,=\,\bigoplus_{c=0}^{r_2}\, (r_1-c)\,\otimes^\prime\, (r_2-c)$$
where $c$ corresponds to the number of contractions between the symmetric traceless tensors.

For the sake of conciseness, we simply explain in CFT language
the idea of the proof: the summation index $l$ in the lemma corresponds to the number of d'Alembertian, the index $m$
to the number of traceless partial derivatives which are not contracted with the current
while the index $n$ corresponds to the number of remaining free indices of the current.
Therefore $s-n$ is equal to the number of
divergences, hence $n$ cannot be higher than $s$ nor smaller than $s-t+1$ (since the $t$-th divergences of the current vanishes).
}

\noindent\textbf{Example:} In the particular case $d=3$, the lemma
\bref{partconscur} and the equality \eqref{partconscurrmod} imply
that the irreducible $\mathfrak{o}(3,2)$-module ${\cal
D}\left(2+s-t,s\right)$ of the $d=3$ partially conserved current of
spin $s\leqslant t$ and depth $t$ decomposes as
\begin{equation}
 {\cal D}\left(2+s-t,s\right)=\bigoplus_{k=0}^{t-1}\,\bigoplus_{m=0}^\infty\,\bigoplus_{p=s+1-t+k}^\infty {\cal Y}\big(m+p+t-2k,p\big)= \frac{{\cal V}\left(2+s-t,s\right)}{{\cal V}\left(2+s,s-t\right)}
\end{equation}
and thus
has character given by the rational function
\begin{eqnarray}\label{charformpcc}
&&\chi\left({\cal D}\left(2+s-t,s\right)\right)=\sum\limits_{k=0}^{t-1}\,\sum\limits_{m=0}^\infty\,\sum\limits_{p=s+1-t+k}^\infty \alpha^{2(m+p+t-2k)} \chi_p(\beta)\nonumber\\
&=&\frac1{(\alpha-\alpha^{-1})(\beta-\beta^{-1})}%\sum\limits_{k=0}^{t-1}\,
\left[\frac{(\alpha\beta)^{2(s-t)+1}\big((\alpha\beta)^{2t}-1\big)}{\alpha\beta-(\alpha\beta)^{-1}}
-\frac{(\alpha/\beta)^{2(s-t)+1}\big((\alpha/\beta)^{2t}-1\big)}{\frac{\alpha}{\beta}-\frac{\beta}{\alpha}}
\right]\nonumber\\
&=&
\chi\left({\cal V}\left(2+s-t,s\right)\right)-\chi\left({\cal V}\left(2+s,s-t\right)\right)
\nonumber
\end{eqnarray}

Endowed with the lemmas \bref{confscal} and \bref{partconscur}, the
generalized Flato--Fronsdal theorem follows by direct computation: On
one hand, the lemma \bref{confscal} together with the rules
\eqref{sumprod} imply that the tensor product of two conformal
scalars of order $2\ell$ is equal to
\begin{eqnarray}
&& {\cal D}\left(\frac{d-2\ell}2,0\right)\otimes {\cal D}\left(\frac{d-2\ell}2,0\right)=\nonumber\\
&&\qquad
=\bigoplus_{m,n=0}^\infty\bigoplus_{q,r=0}^{\ell-1} {\cal Y}\left(d-2\ell+m+n+2q+2r,m\otimes n\right)
\nonumber\\
&&\qquad
=\bigoplus_{l,m,n=0}^\infty\bigoplus_{q,r=0}^{\ell-1} {\cal Y}\left(d-2\ell+2l+m+n+2q+2r,m\otimes^\prime n\right)
\,.\label{lhsgff}
\end{eqnarray}
On the other hand, the right-hand-side of the generalized Flato--Fronsdal theorem is equal to
\begin{eqnarray}
&&\bigoplus\limits_{s=0}^\infty\bigoplus\limits_{k=1}^{\ell}{\cal D}(d+s-2k,s)=\label{rhsgff}\\
&&=\bigoplus\limits_{s=0}^\infty\,\,\bigoplus\limits_{k=1}^{\ell}\,\,\bigoplus_{l',m=0}^\infty\,\bigoplus_{n=\max\{0,s-2k+2\}}^{s}
{\cal Y}(d+2s+2l'-2k+m-n,m\otimes^\prime n)\,.\nonumber
\end{eqnarray}
Notice that the multiplicites are finite in the decompositions \eqref{lhsgff} and \eqref{rhsgff} (indeed the conformal weight grows with
respect to all summation indices with infinite ranges).
The equality between \eqref{lhsgff} and \eqref{rhsgff} follows by exchanging the three summation indices $s$, $k$ and $l'$ for the three indices $l$, $q$ and $r$ via the relation $l=l'-k+\ell$, $q+r=s-n$ and $|q-r|=2k-2$. The index $l$ must then ranges from zero to infinity (since $k\leqslant \ell$ and $l'$ ranges from zero to infinity) and the indices $q$ and $r$ must range from $0$ to $\ell-1$ (since $s-n$ and $2k-2$ range from $0$ to $2\ell-2$) as they should.

\noindent\textbf{Example:} For $d=3$, one should simply check the
following algebraic equality of rational functions:
\begin{eqnarray}
&&\left(\,\chi\left({\cal
D}(\frac32-\ell,0)\right)\,\right)^2=\sum\limits_{s=0}^\infty\sum\limits_{k=1}^{\ell}\chi\left({\cal
D}(3+s-2k,s)\right)\nonumber\\
&&=
\sum\limits_{k=1}^{\ell}\left[
\sum\limits_{s=0}^\infty\chi\left({\cal V}(3+s-2k,s)\right) -
\sum\limits_{s=2k-1}^\infty\chi\left({\cal V}(2+s,s+1-2k)\right)
\right]\,,
\end{eqnarray}
where \eqref{charformpcc} has been used. The $\mathfrak{o}(3,2)$
character formulae \eqref{charVerma} and \eqref{charformcs} should
be further used to verify the equality.

\section{Boundary values and global symmetries}

\subsection{Boundary values in terms of gauge-invariant quantities: an example}
\label{sec:GI}
%\subsection{Spin $1$ example}
The free spin $s$ field on constant curvature space can be described either in terms of the potentials (Fronsdal fields) or in terms of the curvatures. Strictly speaking these two formulations are not completely equivalent as local gauge theories because in one case there is a genuine gauge invariance while in the other there is not. It is instructive to perform the near-boundary analysis for the equations of motion in terms of the curvatures.

For simplicity we restrict ourselves to the first nontrivial case of $s=1$. In this case it is useful to encode the curvature in terms of the ambient generating function depending on auxiliary anti-commuting variables $\theta^A$ which can be identified with basis De Rham differentials $dX^A$. Consider the algebra $\mathfrak{osp}(2|2)$ formed by
\begin{equation}
\begin{gathered}
\label{F-const-2}
 \Box=\half \d_X\cdot \d_X, \qquad \bar\Box=\half X\cdot X\,, \qquad D=\Theta\cdot \d_X \,,\\
 \bar D =\Theta \cdot X\,, \qquad \Delta =\d_\Theta \cdot \d_X\,, \qquad \bar\Delta =X \cdot \d_\Theta\\
  H_+=d+2+X\cdot \d_X-\Theta \cdot \d_\Theta \,, \qquad H_-=X\cdot \d_X +\Theta \cdot \d_\Theta
  %\bar D =\Theta \cdot X\,, \qquad \bar D =\Theta \cdot X\,, \qquad \bar D =\Theta \cdot X\,, \qquad %\bar D =\Theta \cdot X\,, \qquad
\end{gathered}
\end{equation}
The AdS Maxwell equations $df=0$ and $\nabla_\mu f^{\mu\nu}=0$ on AdS can be written in terms of the ambient space 2-form $F=F_{AB}(X)\Theta^A\Theta^B$ as
\begin{equation}
\label{F-const}
 DF=0\,, \quad \Delta F=0\,, \quad \Box F=0\,,\quad X\cdot \d_\Theta F=0\,, \quad H_- F =0\,.
\end{equation}
Note that the last equation implies that the homogeneity degree is $-2$ which suggests that
this formulation is adapted to describing shadow-type boundary values. Indeed, the analysis of~\cite{Bekaert:2012vt} can be applied to the system~\eqref{F-const-2} almost without alternations
and shows that the boundary value is given by closed $2$-form in $d$ -dimensions. It is unconstrained
and in case of even $d$ is subject to FT equations written in terms of $F$.
More precisely, repeating the analysis of~\cite{Bekaert:2012vt} in this case (this is almost straightforward)
one finds that the boundary value $f=f_{ab}\theta^a \theta^b$ is subject to $(\d_x \cdot \d_x)^{\frac{d-4}{2}}(\d_\theta\cdot \d_x)f=0$ along with $\theta \cdot \d_x f=0$, \textit{i.e.} the spin 1 conformal equations in terms of the curvature.

To make contact with the formulation in terms of the potential let us write the Maxwell system in terms of $A=\phi_B\Theta^B$ using the ambient space. The relevant system of constraints is the same but the only difference is that the constraint $D$ is now a gauge generator. More precisley,
\begin{equation}
\Box A=\Delta A=X\cdot \d_\Theta A= H_-A=0\,, \qquad A \sim A+D\lambda
\end{equation}
where $\lambda$ satisfies the same constraints as $A$. This system is just a rewriting of~\eqref{pmassl} with $t=1$
and $s=1$. It is clear that constraints~\eqref{F-const} for $F$ can be simply obtained from the above ones by taking $F=DA$.

Let us now turn to the formulation adapted to the current type asymptotic conditions. Let us recall first how this is done in terms of potentials. The system is determined by the constraints proposed in~\cite{Bekaert:2012vt}, which we here write for $s=1$ case and in terms of $A=\phi_A\Theta^A$.
The system reads
\begin{equation}
\Box A=\Delta A= (H_+-2)A=0\,, \qquad A \sim A+D\lambda + (X\cdot\Theta)\,\epsilon\,,
\end{equation}
where $\lambda$ and $\epsilon$ satisfy the gauge parameter version of the first 3 constraints. Namely,
\begin{equation}
 \Box \epsilon=(H_+-2)\epsilon=0\,,\qquad \Box \lambda+\epsilon=(H_+-4)\lambda=0\,.
\end{equation} 
Note that $\Delta\lambda=\Delta\epsilon=0$ identically because $\lambda,\epsilon$ are $\Theta$-independent.
It was shown in~\cite{Bekaert:2012vt} that this system determines the current-type boundary value.

For the present purposes it is useful to identify the following partial gauge of the above system. Namely, the gauge where $X\cdot \d_\Theta A=0$. The residual gauge transformations
are
\begin{equation}
\label{A-rgs}
 A\sim A+(d-2)(X\cdot\Theta)\,  \epsilon +D ((X\cdot X)\,\epsilon)
\end{equation}
so that only one independent parameter remains. Notice the system can be seen to become St\"ueckelberg
on the hypercone. Indeed, when $X^2=0$ only vanishing $\epsilon$ generates trivial gauge transformations
so that the space of reducibilities is empty and one concludes there are no genuine (non-St\"ueckelberg) gauge symmetries. Let us finally spell out explicitly the complete set of constraints which now involves partial gauge condition:
\begin{equation}
\label{Apg-const}
 \Box A=\Delta A= (H_+-2)A=X\cdot\d_\Theta A= 0
\end{equation}

Starting from this partially gauge fixed formulation it is easy to construct the formulation in terms of curvature. Indeed, introducing  $F=DA$ one arrives at the following system
\begin{equation}
 DF=\Box F=\Delta F=H_+F=0\,,\qquad F\sim F+(X\cdot\Theta)\,  \lambda
\end{equation}
where $\lambda$ is related to the above epsilon as $\lambda=\frac{1}{d-2}D 
\epsilon$ and is subject to the same constraints as $F$. This gives the 
description of the Maxwell equations in the form adapted to studying boundary 
values with current-type asymptotic. It is easy to get back the description in 
terms of the partially gauge fixed potential by introducing $A=X\cdot \d_\Theta 
F$. One then finds that constraints~\eqref{Apg-const} and gauge transformations 
\eqref{A-rgs} immediately follow from those for $F$. Finally, it was shown 
in~\cite{Bekaert:2012vt} that the boundary value of such $A$ can be identified 
with the conserved boundary current. In terms of $F$ the boundary conserved 
current is directly related to the pullback of a 1-form $X^A F_{AB}$ to the 
hypercone.

Note that taking into account the constraints and the gauge symmetries the map 
relating $F$ and $A$ is one-to-one. The relation is more subtle as far as gauge 
parameters are concerned because strictly speaking the formulation in terms of 
$A$ is a genuine gauge theory (when considered on AdS) while that based on $F$ 
is not because the gauge equivalence for $F$ is of St\"ueckelberg nature.

\subsection{Boundary values of the Minkowski scalar field: a toy model}\label{sec:Minkowski}\label{sub:flat}

As an example of different kind let us apply the parent technique to identify
boundary data for the scalar field on Minkowski space, with the boundary being
the space if initial data $x^0\equiv t=const$.

The parent form of the Klein--Gordon field in Minkowski spacetime is described by the BRST operator
\cite{\BGST}
\begin{equation}
\brst=dx^a\dl{x^a}-dx^a \dl{y^a}+c\left(\dl{y^a}\dl{y_a}-m^2\right)
\end{equation}
acting on the space of ``states'' $\Psi(x,dx|y,c)$ where $dx^A$ and $c$ are anticommuting ghosts.

Let us separate, say, the time variable $x^0\equiv t$ and pullback the system to a spatial slice $t=\const$. It just ammounts to dropping $t,dx^0$. The BRST operator is then given by
\begin{equation}
\brst_{0}\,=\,dx^i\dl{x^i}\,-\,dx^i \dl{y^i}\,+\,c\,\left(\dl{y^i}\dl{y_i}-\left(\dl{y^0}\right)^2-m^2\right)\,.
\end{equation}
where the space components of $y^a$ have been denoted by
$y^i\equiv \vec y$ ($i=1,2,\ldots,d-1$).

For this toy model the analysis is very simple because there is no gauge symmetries. Equations of motion
read as
\begin{equation}
 (\dl{x^i}- \dl{y^i})\Psi(x|y)=0\,, \qquad \left(\dl{y^i}\dl{y_i}-\left(\dl{y^0}\right)^2-m^2\right)\Psi(x|y)=0\,.
\end{equation}
The general solution of the second constraint is parametrized in terms of two $y^0$-independent functions $q_y(x,\vec y)$ and $p_y(x,\vec y)$. Indeed, in the space of formal series in $y$-variables $q_y(\vec y)+y^0p_y(\vec y)$
can be uniquley completed to $\phi(\vec y ,y^0)\,=\,q_y\,+\,y^0\, p_y\,+\,{\cal O}\left(\left(y^0\right)^2\right)$
satisfying the second constraint. Finally, taking into account
the first equation one conculdes that the general solution of both constraints is parametrized by two
smooth functions $q(x^i)$ and $p(x^i)$. So that the boundary theory is a theory of two unconsrained fields
$q(x^i)$ and $p(x^i)$.

This theory describes precisely the Hamiltonian phase space of the Klein-Gordon field. This is in agreement with the general statement that the parent formulation of a gauge theory naturally induces the Hamiltonian formulation when reduced to the initial data surface~\cite{Grigoriev:2010ic,Grigoriev:2012xg} (see also~\cite{Barnich:2010sw}). Of course the procedure of~\cite{Grigoriev:2010ic,Grigoriev:2012xg} also keeps track of the action and the symplectic structure it induces while in the above toy model we have only concentrated on the equations of motion.

\subsection{Higher depth shadows and Killing tensors: equivalent descriptions}
\label{sec:shadows-app}
We now use parent formulation in order to define higher depth shadow fields as fields on a conformal space in a manifest way. To this end we construct a parent BRST formulation based on the constrained system~\eqref{shad-amb}. The BRST operator is given by
\begin{equation}
 \Omega^{P}=\nabla+\bar\Omega\,, \qquad 
\end{equation} 
and it acts on the states $\Psi=\Psi(x^a|\theta^a,Y,P,\text{ghosts})$ which are defined on the conformal space with
coordinates $x^a$. Here $\nabla$ is a flat covariant derivative given by the same fomula~\eqref{covder} as in the AdS case but constructed using the flat conformal connection and the conformal version of the compensator field (i.e. with $V^2=0$) and $\bar\Omega$ is the BRST operator implementing the fiber version of the constraints~\eqref{shad-amb}. More precisely,
\begin{equation}
 \bar\Omega=
 %c_\Box
 \bar\Box
 \dl {b_{-1}}+
 %+c_S
 \bar S
 \dl{b_0} +
 %+ c_T
 \Bar T
 \dl{b_1}+ (\sd)^t\dl{b} +c_0 U_-+ c_{-1}\bsd+\ldots
\end{equation}
where dots denote ghost terms involving also the structure operators of the gauge algebra  (for $t=1$ these are  structure constants).  Here the constraints act on functions in $Y$ and $P$ in the ``twisted'' realization
i.e. $X^A$ and $\dl{X^A}$ are represented as $Y^A+V^A$ and $\dl{Y^A}$ with the conformal choice of the compensator $V$, \textit{i.e.} $V^2=0$. 

In order to see that the above system is indeed just a parent reformulation of the constrained system~\eqref{shad-amb}
we note that taking as a degree minus the homogeneity in ghosts $b_{-1},b_0,b_{1}$ the first three terms in $\bar\Omega$ form the lowest degree $-1$ part of $\Omega^P$. Its cohomology can be chosen $b$-idependent and can be identified with the quotient space of all elements modulo the ideal generated by $\bar\Box,\bar T,\bar S$. Reducing to the cohomology gives precisely the parent version of the constrained system~\eqref{shad-amb}.

Let us show that $\Omega^P$ indeed describes higher-depth shadow fields on the conformal space. To this end we
work in the frame where $V^+=1,V^-=V^a=0$ and take as a degree homogeneity in $b_{-1},b_0,Y^+,P^+$.
The lowest degree term in $\Omega_P$ is 
\begin{equation}
%\begin{multline}
\Omega^P_{-1}=
 %+c_S
(p_a y^a+P^-)
 \dl{b_0} 
 +
 %+ c_T
 (y^ay_a+2Y^-) \dl{b_{-1}}+c_0\dl{Y^+}+c_{-1}\dl{P^+}\,.
%\end{multline}
\end{equation} 
Its cohomology can be identified with $b_0,b_{-1},c_{0},c_{-1}$ and $Y^+,Y^-,P^+,P^-$-independent elements.
Because the cohomology is vanishing in nonzero degree the reduced system is simply described by $\Omega^P$
restricted to the cohomology. Using a Cartesian coordinates $x^a$ and the adapted local frame the reduced operator takes the form
\begin{equation}
 \Omega^P_{red}=\theta^a(\dl{x^a}-\dl{y^a})+(p\cdot \d_y)^t\dl{b}+p^2\dl{b_1}\,.
\end{equation} 
In its turn, this is a straitforward parent rewriting of the BRST operator $\Omega_{red}=(p\cdot \d_x)^t\dl{b}+p^2\dl{b_1}$
defined on $\Psi(x,p,b,b_1)$. It obviosuly encodes trivial equations and the following gauge transformations
\begin{equation}
 \delta \phi=\Omega_{red} \Lambda=(p\cdot \d_x)^t \lambda+p^2\chi\,,  \qquad \Lambda=b \lambda+b_1 \chi
\end{equation} 
One concludes that the reduced system precisely describes higher-depth shadow 
fields. Furthermore, (higher-depth) Killing tensors are 1:1 with global reducibility parameters which are in turn 1:1
with BRST-cohomology elements at ghost degree $-1$. Cohomology elements of $\Omega_{red}$ at degree $-1$ are by definition
higher-depth Killing tensors. Because cohomology does not depend on the formulation,  
this gives the ambient characterization of higher-depth Killing tensors given in~\eqref{Ckill-dual}.

Our next aim is to show that higher depth conformall Killing tensors can be 
equivalently described by~\eqref{Ckill}. To this end we compute the cohomology of 
the BRST operator $\Omega^P$ at ghost degree $-1$. Taking as a degree 
homogeneity in $b_{-1},b_0,b_1$ the lowest degree term in $\Omega^P$ is a sum of 
first three terms in $\bar\Omega$. Its cohomology can be assumed 
$b_{-1},b_0,b_1$-independent and hence a representative of $\Omega^P$-cohomology 
can also be assumed $b_{-1},b_0,b_1$-independent. Such a representative 
$\varepsilon$ of ghost degree $-1$ is necesarily linear in $b$ and 
$\theta^a,c_{-1},c_0$-independent so that $\varepsilon=b \epsilon(x|p,y)$. The 
cocycle condition then implies $\bsd\epsilon =(\sd)^t\epsilon=0$ and hence 
$\epsilon$ is polynomial in $Y$.  By adding a coboundary a polynomial cocycle 
$\varepsilon$ can always be assumed totally traceless i.e. 
$T\varepsilon=S\varepsilon=\Box \varepsilon$. Indeed, in the space of 
polynomials the cohomology of the sum of first three terms in $\bar\Omega$ can 
be identified with totaly traceless $b_{-1},b_0,b_1$-elements. Furthermore, a 
direct analysys show that such $\epsilon$ can not become trivial in the space of 
formal series. Finally, beeing polynomial such $\epsilon$ are one-to-one with 
$x$-independent ones because $x$-dependence is uniquely detremined by $\nabla 
\epsilon=0$. The later relation is implied by $\Omega^P\varepsilon=0$ because 
$\varepsilon$ is $\theta$-independent. In this way we have shown that 
higher-depth Killing tensors can equivalently be represented by~\eqref{Ckill}.

\subsection{Global symmetries of higher-order singletons: the proof}
\label{sec:symm-proof}

The proof can be given in the first quantized terms following the ideas and the technique of~\cite{Bekaert:2009fg}. The global symmetries of the equation $\Box_0^l\phi=0$ can be seen as observbles of the constrained system with the only constraint $\Box_0^l$. These are conveniently represented as the ghost degree zero BRST cohomology of the adjoin action $D=\qcommut{\Omega}{\cdot}$ of the BRST operator
\begin{equation}
 \Omega=c (p^2)^l
\end{equation} 
where we have switched to the star-product conventions. Indeed, the cocycle condition can be written as
\begin{equation}
 \qcommut{(p^2)^l}{A}+B*(p^2)^l =0
\end{equation} 
And the co-boundary condition amounts to
\begin{equation}
 A\sim A+C* (p^2)^l\,.
\end{equation} 
As we have already seen using the ambient space this constrained system can be equivalently represented through the BRST operator implementing the following constraints 
\begin{equation}
 (X^2)^l\,, \qquad X\cdot P\,, \qquad P\cdot P 
\end{equation} 
More precisely, the Weyl symbol of the BRST operator is given in~\eqref{HOS-BRST-Weyl}. Its adjoint action reads as
\begin{multline}
\label{adj-omega}
 D=\half{\left(\half{P^2}\right)^l}\dl{b_1}+(X\cdot P)\dl{b_0}+\half{X^2}\dl{b_{-1}}+
 \\
 +c_1\qcommut{(\half {P^2})^l}{\cdot}+
 c_0 U_-+c_{-1}\bsd +\ldots\,,
\end{multline} 
where dots denote terms of higher order in ghosts and ghost derivatives.

Using a suitable degree the terms in the first line form a lowest degree term 
$\delta$ and one can reduce the cohomology problem to its cohomology.  It is 
enough to compute $\delta$ cohomology in an irreducible component of $\mathfrak{sp}(4)$ 
module of polynomials in $X,P$ (tensored with ghosts). According to~\cite{\BGST} 
such a module is a generalized Weyl module freely generated by 
$X^2,P^2,X\cdot P$ from some irreducible $\mathfrak{sp}(2)$-module.
In this component one can treat $\bar \Box=\half X^2$, $\bar S= XP$, $\bar T=\half P^2$
as formal commuting variables. 
{It follows the $\delta$-cohomology can be identified with 
$b_\alpha$-independent elements of the form 
$\psi=\psi_0+\bar T\psi_1+\ldots {\bar T}^{l-1} \psi_{l-1}$
and hence a cocycle $\psi$ can be assumed to 
have this form.} The reduced cohomology problem says that 
$(D \psi)|_{{\bar T}^l=\bar\Box=\bar S=0}=0$.
As $\psi$ is ghost independent (thanks to $\gh{\psi}=0$) this gives 
\begin{equation}
 \bsd \psi_k=0\,, \qquad U_-\psi_k=-2k
\end{equation} 
for a homogeneous $\psi={\bar T}^k \psi_k$.

Consider an irreducible $\mathfrak{sp}(2)$ submodule in the module of degree $k$ polynomials in 
$\bar \Box, \bar S,\bar T$, which grows from highets weight vector  ${\bar T}^k$.
It is clearly $2k+1$-dimensional and as a basis it is useful to take $A_\alpha$ $\alpha=0\ldots 2k$ such that $A_0={\bar T}^k$ and $\sd A^\alpha=A^{\alpha-1}$. Note that $A^{2k}$ is proportional to ${\bar \Box}^k$. Element ${\bar T}^k \psi_k$ has a completion $\Phi$ by $\bar \Box,\bar S$-dependent elements of total (in $\bar \Box, \bar S,\bar T$) degree $k$ such that $\sd \Phi=0$. Indeed, taking
\begin{equation}
 \Phi=\sum_{\alpha=0}^{2k} (-1)^{\alpha} A^\alpha (\sd)^\alpha \psi_k
\end{equation} 
one finds that $\sd\Phi=0$. Indeed, $\sd A^0=0$ by consruction and $(\sd)^{2k+1}\psi_k=0$ thanks to $U_- \psi_k=2k$. 
The completion is unique if in addition the space has been quotient with respect to $(\bar S)^2-4\bar \Box \bar T$.  The constructed $\Phi$ is of degree $k$ in $\bar \Box, \bar S,\bar T$
and satisfies $U_-\Phi=\bsd\Phi=0$ so that $\sd \Phi=0$ as well. One then concludes that the cohomlogy is one-to-one with rectangular and $\ell$-traceless elements which are in addition anihilated by $S^2-4T\Box$.

\pagebreak

%%%%%%%%%%%%%%%%%%%%%%%%%%%%%%%%%%%%%%%%%%%%%%%%%%%%%%%%%%%%%%%%%%%%%%%%%%%%%%%%%%%%%%%%%%%%%%%%%%%%%%
%\small{
%\bibliography{/home/maxim/Documents/HSmaster}}
\renewcommand{\refname}{References\\[5pt]}
\small{
%\bibliography{/home/maxim/Documents/HSmaster}}
\addtolength{\baselineskip}{-7pt}
%\parsep=-8pt
%\itemsep=-4pt
\addtolength{\itemsep}{-10pt}
%\addtolength{\baselinestretch}{1pt}{1pt}

%\addtolength{\parskip}{-10pt}
%\bibliography{HSmaster}}

\providecommand{\href}[2]{#2}\begingroup\raggedright\endgroup

\end{document}